\documentclass{aa}  
\usepackage[]{graphicx}
\usepackage{hyperref}
\usepackage{txfonts}
\usepackage{float}

\usepackage{amsmath}
\usepackage[ ]{natbib} 
\usepackage{multirow}

\bibpunct{(}{)}{;}{a}{}{,} 

\newcommand{\xmm}{\textit{XMM-Newton} }
\newcommand{\nustar}{\textit{NuSTAR} }

\newcommand{\swift}{\textit{Swift/XRT} }

\newcommand{\source}{\rm{MXB 1659-298} }

\begin{document} 

   \title{Broadband spectral analysis of MXB 1659-298 in its soft and hard state}
%
%

\author{R. Iaria\inst{1}\and S. M. Mazzola\inst{1}\and
  T. Bassi\inst{1,2,3}\and A. F. Gambino\inst{1}\and A. Marino\inst{1,2,3}\and
  T. Di Salvo\inst{1}\and A. Sanna\inst{4}\and  A. Riggio\inst{4}\and
L. Burderi\inst{4}\and N. D'Amico\inst{4,5}}

\institute{Universit\`a degli Studi di Palermo, Dipartimento di Fisica
  e Chimica, via Archirafi 36 - 90123 Palermo,
  Italy\label{inst1}\\\email{rosario.iaria@unipa.it} \and 
INAF -- Istituto di Astrofisica Spaziale e Fisica Cosmica di Palermo,
Via Ugo La Malfa 153, I-90146 Palermo, Italy\label{inst2}\and  
IRAP, Universit\'{e} de Toulouse, CNRS, UPS, CNES, Toulouse, France\label{inst3}\and 
Universit\`a
  degli Studi di Cagliari, Dipartimento di Fisica, SP
  Monserrato-Sestu, KM 0.7, 09042 Monserrato, Italy\label{inst4} 
\and INAF, Osservatorio
  Astronomico di Cagliari, Via della Scienza 5, I-09047 Selargius
  (CA), Italy\label{inst5}
           }

          
            
%


 %
  \abstract
   {The X-ray transient eclipsing source \source went into outburst in
     1999 and 2015. During these two outbursts the source was observed
     by \xmm\!, \nustar, and \swift. }
   {Using these observations, we studied the broadband spectrum
 of the source to constrain the continuum components and to verify whether it had a reflection component, as is observed in other X-ray
eclipsing transient sources. }
   {We combined the available spectra to study the soft and hard state
   of the source in   the  0.45-55 keV energy range. }
   {  We report a reflection component in  the soft and
    hard state. The direct emission in the soft state can be modeled with a thermal
     component originating from the inner accretion disk plus a
     Comptonized component associated with an optically thick corona
     surrounding the neutron star. On the other hand, the direct emission in the hard
     state   is described only by a
   Comptonized component with a temperature higher than 130 keV; this
   component is 
   associated with an optically thin corona. We observed 
        narrow absorption  lines from highly ionized
       ions of oxygen, neon, and iron in the soft spectral state. We investigated where the narrow absorption lines form in the ionized absorber. The
     equivalent hydrogen column density associated with the absorber
     is close to $6 \times 10^{23}$ cm$^{-2}$ and  $1.3 \times
     10^{23}$ cm$^{-2}$  in the soft and hard state, respectively.   }
   {}


\maketitle

\section{Introduction}
\label{sec:Intro}

Low-mass X-ray binaries hosting neutron stars (hereafter NS-LMXBs) are
binary systems in which a weakly magnetized neutron star (NS) accretes
matter from a low-mass ($<1$ M$_{\odot}$) companion star through 
Roche-lobe overflow.  
 The NS-LMXBs display several spectral
states and hysteresis patterns  similar to that observed in X-ray binary systems harboring a black hole (BH) as compact object, as shown by \cite{Munoz_14}.  The hysteresis pattern  highlights  the soft, hard, and intermediate states. This
suggests  a similarity between these two classes of X-ray sources.
 The  NS-LMXB spectrum in the soft state
shows a dominant soft emission associated with a blackbody
or a disk-blackbody component with a temperature between 0.5 and 2 keV 
and a harder (3-5 keV) saturated Comptonized
component with a high optical depth. The thermal component  
 originates from the NS surface and/or from the innermost region of the accretion disk, while 
 the Comptonized component is associated with  inverse Compton emission that is due to the interaction of the soft photons emitted from the innermost region  with the electrons in 
 the thermal corona surrounding the 
 NS \citep[see, e.g.,][and references therein]{Disalvo_09,
  Iaria_09}.  On the other hand, the hard-state spectrum
can be described by a weak (or absent) thermal component plus a cutoff power-law
component, with a cutoff energy higher than 10 keV that is  produced by inverse Compton scattering of soft photons
in a hot optically thin electron corona \citep[see, e.g.,][and references
therein]{Disalvo_15, Cackett_10}.
\begin{figure*}[ht]
{\includegraphics[width=8.5cm]{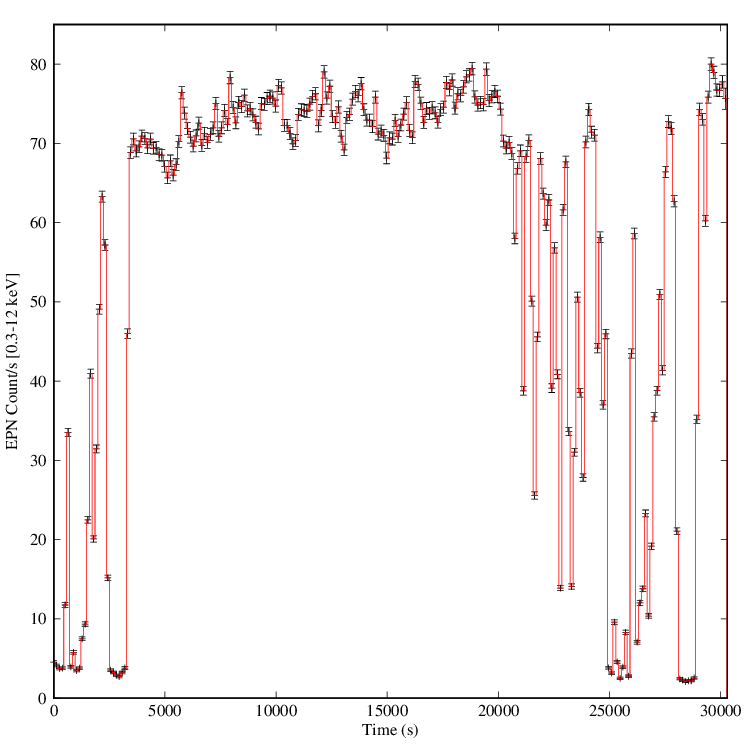}
\includegraphics[width=8.5cm]{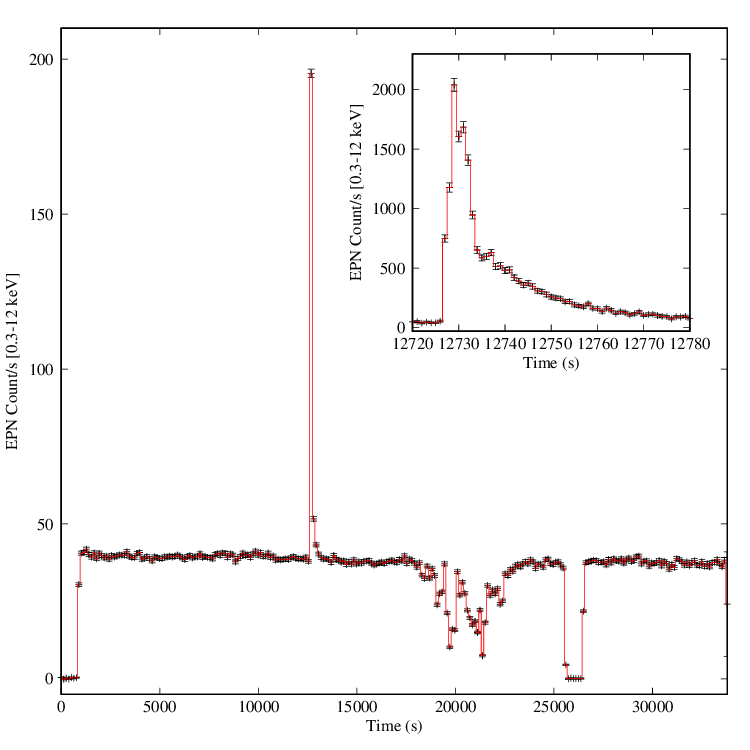}}
\caption{EPN background-subtracted light curve in the 0.3-12 keV energy
  range corresponding to the \xmm observations taken in 2001 (left panel)
  and 2015 (right panel).
The bin time is 128 s. Inset: Shape of the type I X-ray
  burst with a binning time of 1 s.}
\label{fig:epn_light_hard}
\end{figure*}

 Studying the reflection component that originates from direct Compton
scattering of the photons that are emitted by the hot corona with the cold
electrons in the top layers of the inner accretion disk is of particular interest. 
A peculiar feature in the continuum emission is the Compton hump 
 around 20-40 keV.  It can be ascribed to   direct
Compton scattering  and is observed mainly 
when the source is in the hard state because the flux above 20 keV
is higher than in soft state \citep[see, e.g.,][and references
therein]{Egron_13, Miller_13}.
The reflection component also shows   
 broad emission lines (FWHM close to 1 keV) in the Fe-K region, which are
often observed in the spectra of NS-LMXBs with an inclination
angle lower than 60$^{\circ}$ \citep[see,
e.g.,][]{Iaria_16,Iaria_09,Pandel_08,Shapo_09} and
inclination angles between 60$^{\circ}$ and 80$^{\circ}$ \citep[the
so-called dipping and eclipsing sources,
e.g.,][]{Iaria_07,Ponti_15}.  These lines are
identified with the K$_{\alpha}$ transitions of iron at different
ionization states. Compton scattering is not sufficient to explain the
large width measured for these lines
\citep{Sanna_13,Reis_09}.  A possible scenario that allows explaining the large observed broadening is that these lines
originate in the innermost region of the accretion disk, close to the
compact object, where matter reaches Keplerian velocities up to a few
tenths of the speed of light. For this reason the reflection component
is corrected by relativistic Doppler effects and gravitational
redshift \citep{Fabian_89}.

MXB 1659-298 is an eclipsing transient source that shows type I X-ray
bursts. \cite{Iaria_18} and \cite{Chetana_17}, studying the eclipse
arrival times of  the source obtained during
  the outburst that occurred  between 1999 and 2001 and the outburst between 2015
and 2017, suggested that a third body
might orbit the binary system.  \cite{Iaria_18} inferred  an
orbital period of the binary of 7.1161099(3) hr and an orbital period derivative of
$-8.5(1.2) \times 10^{-12}$ s s$^{-1}$; they also inferred an inclination 
angle of $72^{\circ} \pm 3^{\circ} $.  The distance to the source was
estimated to be $9 \pm 2$ and $12 \pm 3$ kpc for a hydrogen-rich and helium-rich
companion star, respectively \citep{Gallo_08}.

Studying the \xmm spectrum of \source\!\!\!, \cite{Sidoli_01}
detected two absorption lines at 6.64 and 6.90
keV associated with \ion{Fe}{xxv} and \ion{Fe}{xxvi} ions,
as well as absorption lines associated with the transitions of 
\ion{O}{viii}  at 0.65, 0.77, 0.81 keV and 
\ion{Ne}{ix} at 1.0 keV, respectively.
Furthermore, the authors detected a broad emission line 
centered at 6.47 keV with a FWHM of 1.4 keV   that can be associated
with a K-shell transition of neutral or weakly ionized iron. 

Recently,  \cite{Sharma_18} analyzed the 0.5-30 keV soft spectral state of the source using {\it Swift/XRT} and \nustar data and fit the continuum emission with a model composed of a thermal component and a Comptonized component with seed photons emitted from the NS surface. The same authors analyzed the hard spectral state using  {\it Swift/XRT} and \nustar spectra in the 0.5-70 keV energy band and adopting a model composed of two thermal components produced at the NS surface and at the innermost region of the disk plus a Comptonized component in which the seed photons are emitted from the NS surface. The same authors also investigated an ionized absorber in the soft and hard spectral states, finding that this is present in the soft state with an ionization parameter log$(\xi)$ close to 4 and a corresponding equivalent column of neutral hydrogen of  $7 \times 10^{23}$ cm$^{-2}$, while the ionized absorber is not statistically significant in the hard spectral state.

\cite{Ponti_19} used {\it Chardra/HETG}   and \nustar data to analyze the 0.8-30 keV soft spectral state and \xmm  data  to analyze the  0.5-10 keV hard spectral state of MXB 1659-298. The HETG spectrum shows the presence a several absorption lines of H-like and He-like ions of Ne, Mg, Si, S, Ar, Ca, Fe, and Ni. The adopted continuum emission is composed of two thermal components corresponding to the emission from the NS surface and the innermost region of the accretion disk plus a Comptonized component associated with the corona surrounding the  NS. The authors also suggest the presence of a broad emission line in the Fe-K region of the spectrum that was interpreted as  a blurred relativistic line. The absorption lines were ascribed to an ionized absorber that has a ionization parameter log$(\xi)$ close to 4.5 and a corresponding equivalent column of neutral hydrogen of  $2.4 \times 10^{23}$ cm$^{-2}$.

 The behavior of MXB 1659-298  is similar to that of the eclipsing transient source AX J1745.6-2901. 
  A broadband study of  AX J1745.6-2901 was performed by \cite{Ponti_15}, who  analyzed the source during its outbursts using {\it XMM-Newton}, \nustar, and {\it Swift/XRT} data. The authors studied the soft and hard spectral states  and found that the persistent emission  is absorbed by a   column density of neutral material of  $N_{H} \simeq 1.9 \times 10^{23}$ cm$^{-2}$ and concluded that  most of the (photoelectric) obscuring column density is due to the interstellar medium.  Moreover,  the same authors  detected Fe K absorption lines in the soft state that disappeared during the hard state. The column densities and turbulent velocities of the absorbing ionized plasma are higher than $10^{23}$ cm$^{-2}$ and  500 km s$^{-1}$, respectively. Finally, they found that the continuum emission in the soft state  requires a relativistic line component with an equivalent width between 80 and 200 eV.

In this paper we report the broadband spectral analysis of
the persistent  X-ray emission of \source during the 1999 and 2015 outbursts 
using {\it XMM-Newton}, \nustar,
and {\it Swift/XRT} data. We analyzed  the X-ray spectra in the soft
  and hard state   and found  that a relativistic reflection component
is necessary to fit the spectra; furthermore, we  observed absorption lines associated with highly ionized iron that
  we modeled using an ionized absorber.  We discuss
  the nature of the ionized absorber in the soft spectral state where the absorption
  lines are more prominent.

\begin{figure*}[!htbp]
\centering
 {  \includegraphics[scale=0.7]{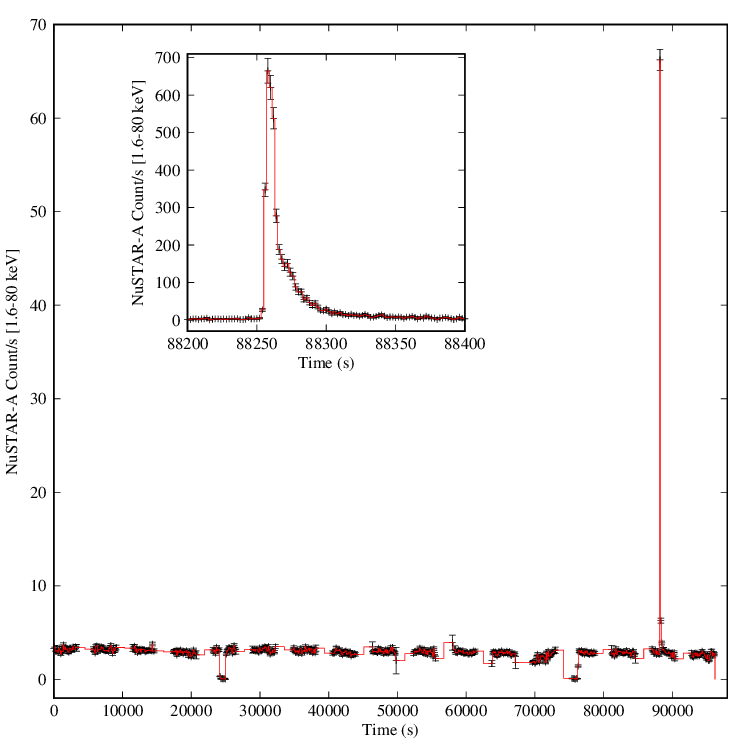}
   \includegraphics[scale=0.7]{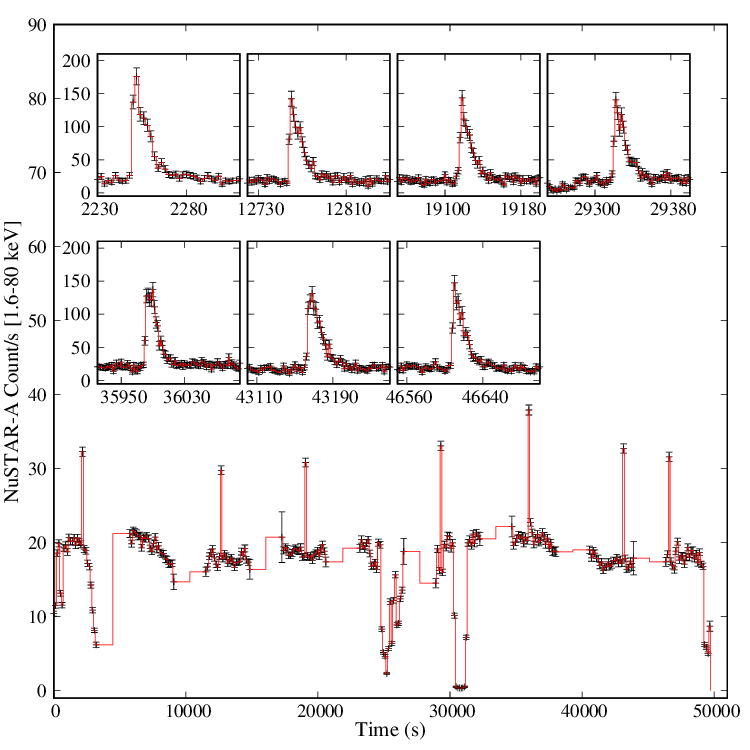}}
 \caption{FPMA light curves obtained from the observations taken in 2015
   (left) and in 2016 (right). The bin time is 128 s. The  details of the observed X-ray type I bursts 
   are shown in the inset panels with a bin time of 1 s.}
              \label{fig:burst}%
    \end{figure*}

\section{Observation and data reduction}

\subsection{\xmm observations}
The \xmm Observatory \citep{jansen_01} observed \source twice.
The first observation (ObsId. 0008610701) was performed on 2001 February 20 between 8:28:27
UTC and 16:2:39 UTC for 31.5 ks, during the outburst that occurred from April 
1999 until September 2001. The source was observed a second time for 42.9 ks between
2015 September 26 19:53:05 UTC and September 27 06:53:47 UTC, during its last
outburst (Obsid.  0748391601).  The first and second \xmm observation have been studied by \cite{Sidoli_01} and \cite{Ponti_19}, respectively.

The PN-type CCD detector of the European Photon Imaging Camera
\citep[EPN,][]{struder_01} was operating in Small Window Imaging (SW) mode  
during the first observation and in Timing mode
during the second observation. The Reflecting Grating Spectrometer \citep[RGS,
two modules,][]{herder_01} was operating in standard spectroscopy mode
during these two observations.  We reprocessed the \xmm data
using the Science Analysis Software (SAS) v16.1.0, obtaining the
calibrated photon event files using the SAS tools {\tt epproc} and
{\tt rgsproc}.  We verified the absence of background flaring   in the EPN
data by extracting the light curve in the 10-12 keV energy range for all
the observations. The EPN events  were 
selected between 0.3 and 12 keV  using only single and double pixel
events  (PATTERN
$\le$ 4)  that were optimally calibrated for spectral analysis (FLAG $=0$). For our aim,
we considered only the RGS and EPN events during the persistent emission, excluding
dips,  eclipses, and type I X-ray bursts. The RGS spectra (hereafter
RGS12 spectra) were obtained by combining the first-order spectra of
RGS1 and RGS2 using the SAS tool {\tt rgscombine} after verifying that
the two spectra were similar to each other.

 The  EPN  light curve of the persistent emission  corresponding to the
observation taken in 2001 was obtained using a circular region with a
radius of 50$\arcsec$ . We found  a count rate higher than 140
counts s$^{-1}$. Because pile-up issues are expected to be
significant at count rates
higher than 25 counts s$^{-1}$ in EPN observations performed in SW mode,  we extracted   the
source and background events using an annular region with inner and
outer radius of 8.1$\arcsec$ and 50$\arcsec$, respectively, 
to mitigate the pile-up issues. We verified the
goodness of our choice using  the SAS tool {\tt epatplot}.  
The 0.3-12 keV  EPN background-subtracted 
light curve is  shown in  the left panel of
Fig. \ref{fig:epn_light_hard}. The light
curve shows two dips, two total eclipses, and a type I
X-ray burst. The count rate during the persistent emission is between
70 and 80 counts s$^{-1}$ and the corresponding exposure time is 12.8
ks. The RGS12 spectrum has an exposure time of 34.6 ks.

 The source events of the second observation were extracted
  from a box region that included the brightest columns of the detector
  and had a width of 127$\arcsec$ (RAWX between 22 and 52). The
  background events were extracted from a box region far away from the
  source with a width of 10$\arcsec$ (RAWX between 5 and 10).  We
  show the EPN background-subtracted light curve in the right panel of
  Fig. \ref{fig:epn_light_hard}.  The light curve shows a dip at 20 ks from the start time, two eclipses (the first at the
start time of the observation and the second 26 ks after the start
time) and a type I X-ray burst at 12.7 ks. We excluded the events
collected later 33.7 ks from the beginning of the observation because
they were affected by an instrument detection failure.  The count rate
during the persistent emission is 40 counts s$^{-1}$; because the EPN
operated in Timing Mode, we can exclude pile-up issues for this
observation. The RGS12 and EPN spectra have an exposure time of 68.7
and 26 ks, respectively. We adopted  energy ranges of 0.4-2 keV for the RGS12 spectra and 0.6-12 keV for the
EPN spectra.

\subsection{\nustar observations}
    During the outburst that occurred between 2015 and 2017, \source was
    observed twice by the Nuclear Spectroscopic Telescope Array
    satellite \citep[\textit{NuSTAR},][]{harrison_13}: the first
    observation (ObsId. 90101013002) was performed between 2015 September 28
    21:51:08 UTC and September 30 00:56:08 UTC for an exposure time of 51.5
    ks, the second observation (ObsId. 90201017002) was carried out
    between 2016 April 21 14:41:08 UTC and April 22 04:56:08 UTC for an exposure
      time of 26.8 ks. The analysis of the observation taken in 2015 was also reported by \cite{Sharma_18}, and the observation taken in  2016 was
      reported by    \cite{Sharma_18} and \cite{Ponti_19}.

    The data were processed using the \nustar Data Analysis Software
    (NuSTAR-DAS) v1.9.3 for the data sets
    taken by the focal plane modules, FPMA and FPMB, and for both
     observations. 
    The source events were extracted from  a circular region
    centered on the source coordinates that had a radius of
    110$\arcsec$. The filtered events, the
    background-subtracted light curves, the spectra, and the  arf and rmf
    files  were created using {\tt nuproducts} tool.

 We show the 1.6-80 keV FPMA background-subtracted light curves for the
first and second {\nustar} observation in
Fig. \ref{fig:burst}.
We identified two dips, two
eclipses, and one type I X-ray burst in the light curve corresponding
to the first observation, while in the
second observation  we observed seven type I X-ray
bursts, one eclipse, one almost complete   dip at 25 ks after
the start time, and part of two dips at the beginning and end of the
observation. The count rate during
the persistent emission is close to 4 count s$^{-1}$ and 20 count
s$^{-1}$ during the first and second observation, respectively.

To produce the FPMA and FPMB spectra associated with the persistent emission,  
we created, using {\tt XSELECT} v2.4d,  the
good time intervals (gti) files  to exclude the type I X-ray bursts, the 
eclipses, and the dips.  Finally, after  verifying the good agreement
between the FPMA and FPMB spectra, we used the {\tt addascaspec} task
to obtain a single combined  spectrum. The exposure times of the combined \nustar
persistent spectrum are 92.8 ks and 44.3 ks for the first and the
second observation, respectively.
We adopted  energy ranges of 3-35 keV and 3-55 keV for the spectra
extracted from the observations taken in 2015 and 2016, respectively.

\subsection{\swift observation}

\source was monitored   by the {\it X-Ray Telescope}
\citep[XRT,][]{burrows05} on board the {\it Swift} Observatory
\citep{gehrels04}.   We analyzed the \swift  observations 
 00034002036 and 00081918001 taken on 2016 April 20 01:47:54 
and 2016 April 21 20:39:01, respectively, that is, 37 hours before and 7 hours
after the beginning of the \nustar    observation taken in 2016. 
The two observations lasted for 803 s and 697 s, respectively,
and were performed in Window Timing (WT) mode\footnote{http://www.swift.ac.uk/analysis/xrt/modes.php}.
 We used the software package
HEASOFT (v.6.20) with the {\it Swift} Calibration Database (CALDB
v.20160609), running the task {\it xrtpipeline}.  

During the first observation,  the roll angle of the
satellite \citep[for more details
see][]{burrows05} changed, which produced two different strips in the image of the source.
For this reason we 
generated two separate event files with the ftool XSELECT (v. 2.4d),
one for each    strip, by extracting the corresponding
products. We extracted the source spectra from a circular region
with a radius of 20 pixel (1 pixel= 2.357$''$) centered on the source
position using the task {\it xrtproducts}. The background was
extracted from a same-size region that was free from the source.  
The two spectra  have exposure times of 669 s and 134 s,
respectively. 
The second observation    was not affected by the change in roll angle, and   we extracted the
source and background spectra  using
the same criteria  described above. 
We verified that the spectral shape was compatible among  the three spectra  and summed them using the FTOOLS {\tt addascaspec}. 
 The combined spectrum has an exposure time of 1.4 ks and it was rebinned to have at least 20 counts
per energy bin so that we could apply the $\chi^2$ statistics. The adopted
spectral energy range is 0.5-9 keV.

\section{Spectral analysis}

 We combined the fits of   the  spectra obtained
from the \xmm and \nustar observations taken in 2015 
because they show a similar flux of 5  mCrab, while the
 spectra obtained from  
the \xmm observation taken in 2001 and the  \nustar and \swift observations taken in 2016   were fit together  because they   show a flux 
close to 30 mCrab 
\citep[see the ASM and MAXI/GSC light curves, shown in Fig. 1 by][]{Iaria_18}.
Hereafter we call  the two combined spectra {\it \textup{low-}} and 
 {\it  \textup{high-flux spectrum}}, respectively.

\begin{table*} [!htbp]
\centering
\scriptsize
\begin{tabular}{llcccccccccc}
\hline
\hline

Model & Component &\multicolumn{2}{c}{Model 1}
  &\multicolumn{2}{c}{Model 2} &\multicolumn{2}{c}{Model
                                 2}
  &\multicolumn{2}{c}{Model 3} \\
   &   &     &   &    &    &   \multicolumn{2}{c}{+Broad Line} & & \\  
  &    & Obs. 2001 & Obs. 2016 & Obs. 2001 & Obs. 2016 & Obs. 2001 & Obs. 2016 & Obs. 2001 & Obs. 2016  \\
\hline

{\sc edge} & E (keV) & \multicolumn{2}{c}{$0.529^{+0.012}_{-0.010} $} & \multicolumn{2}{c}{$0.536^{+0.033}_{-0.003} $ } & \multicolumn{2}{c}{$0.537^{+0.028}_{-0.004} $} & \multicolumn{2}{c}{$0.541^{+0.026}_{-0.011} $} \\
 & $\tau$ & \multicolumn{2}{c}{$0.10 \pm 0.03$} & \multicolumn{2}{c}{$0.06 \pm 0.03$}& \multicolumn{2}{c}{$0.06 \pm 0.02$} & \multicolumn{2}{c}{$0.06 \pm 0.02$} \\\\
{\sc zxipcf} &N$_{\rm H}$ (10$^{22}$ atoms cm$^{-2}$) &
                                                        \multicolumn{2}{c}{-}&
                                                                               \multicolumn{2}{c}{$120
                                                                               \pm
                                                                               20$}
                               & \multicolumn{2}{c}{$62 \pm 12$} & \multicolumn{2}{c}{$60^{+9}_{-14}$} \\
& $f _{\rm IA}$ & \multicolumn{2}{c}{-}&
                               \multicolumn{2}{c}{$0.88\pm0.07$}
  &  \multicolumn{2}{c}{$>0.97$}&  \multicolumn{2}{c}{1 (frozen)}  \\
 & log$(\xi)_{\rm IA}$ & \multicolumn{2}{c}{-}& $4.45^{+0.14}_{-0.09}$ &
                                                                 $4.24\pm0.05$&
                                                                                         $4.366^{+0.048}_{-0.008}$ & $4.11 \pm 0.03$  &  $4.37 \pm 0.04$& $4.14 \pm 0.05$\\\\
 
{\sc TBabs} & N$_{\rm H_{ISM}}$ (10$^{22}$ atoms cm$^{-2}$) &
                                                        \multicolumn{2}{c}{$0.267 \pm 0.007$} & \multicolumn{2}{c}{ $0.282^{+0.008}_{-0.013}$} & \multicolumn{2}{c}{ $0.281 \pm 0.003$}& \multicolumn{2}{c}{ $0.275 \pm 0.010$} \\\\

  {\sc partcov} & $f$ &   \multicolumn{2}{c}{-}&  \multicolumn{2}{c}{$0.88\pm0.07$} &  \multicolumn{2}{c}{$>0.97$}&  \multicolumn{2}{c}{1 (frozen)}  \\
  {\sc cabs} &N$_{\rm H}$ (10$^{22}$ atoms cm$^{-2}$) &
                                                        \multicolumn{2}{c}{-}& \multicolumn{2}{c}{$120\pm 20$} & \multicolumn{2}{c}{$62\pm12$} & \multicolumn{2}{c}{$60^{+9}_{-14}$} \\\\

{\sc diskbb} & kT$_{\rm in}$ (keV) & \multicolumn{2}{c}{$0.437^{+0.009}_{-0.024} $} & \multicolumn{2}{c}{ $0.316^{+0.028}_{-0.008}  $} & \multicolumn{2}{c}{ $0.314 \pm 0.011 $}&\multicolumn{2}{c}{$0.30\pm0.02 $} \\
 & R$_{\rm disk} \sqrt{\cos \theta}$ (km) &
                                            \multicolumn{2}{c}{$17.8^{+1.8}_{-1.1}$}
  & \multicolumn{2}{c}{ $45^{+5}_{-10}$} & \multicolumn{2}{c}{ $39 \pm
                                           3$} &
                                                 \multicolumn{2}{c}{$39\pm4 $}  \\\\

{\sc gauss} &   E (keV) & \multicolumn{2}{c}{-}&\multicolumn{2}{c}{-}&
                                                                       \multicolumn{2}{c}{$6.60
                                                                       \pm
                                                                       0.10
                                                                       $}
  &\multicolumn{2}{c}{-}\\
 &  $\sigma$ (keV)&
                    \multicolumn{2}{c}{-}&\multicolumn{2}{c}{-}&\multicolumn{2}{c}{$0.69
                                                                 \pm
                                                                 0.08$}
  &\multicolumn{2}{c}{-}\\
& Norm. ($10^{-3}$ ph. cm$^{-2}$ s$^{-1}$) &
                                          \multicolumn{2}{c}{-}&\multicolumn{2}{c}{-}&
                                                                                       \multicolumn{2}{c}{$1.10
                                                                                       \pm
                                                                                       0.10$}&\multicolumn{2}{c}{-}\\\\
{\sc rdblur} & Betor10 & \multicolumn{2}{c}{-}& \multicolumn{2}{c}{-}& \multicolumn{2}{c}{-}& \multicolumn{2}{c}{$<-1.4$} \\
 & R$_{\rm in}$ $\rm (GM/c^2)$ & \multicolumn{2}{c}{-}& \multicolumn{2}{c}{-} & \multicolumn{2}{c}{-} & \multicolumn{2}{c}{$60^{+60}_{-30}$} \\
 & R$_{\rm out}$ $\rm (GM/c^2)$ & \multicolumn{2}{c}{-}& \multicolumn{2}{c}{-} & \multicolumn{2}{c}{-} & \multicolumn{2}{c}{$2800$ (frozen)} \\
 & $\theta$ (deg) & \multicolumn{2}{c}{-}& \multicolumn{2}{c}{-}  & \multicolumn{2}{c}{-}   & \multicolumn{2}{c}{ 72 (frozen)} \\\\

{\sc rfxconv} & rel$_{\rm refl}$& \multicolumn{2}{c}{-}& \multicolumn{2}{c}{-} & \multicolumn{2}{c}{-} & \multicolumn{2}{c}{$0.22^{+0.12}_{-0.05}$} \\
 & Fe$_{\rm abund}$ & \multicolumn{2}{c}{-}& \multicolumn{2}{c}{-} & \multicolumn{2}{c}{-} & \multicolumn{2}{c}{1 (frozen)} \\
 & $\cos \theta$ & \multicolumn{2}{c}{-}& \multicolumn{2}{c}{-} & \multicolumn{2}{c}{-} & \multicolumn{2}{c}{0.309 (frozen)}\\
 & log($\xi$) & \multicolumn{2}{c}{-}& \multicolumn{2}{c}{-} & \multicolumn{2}{c}{-} & \multicolumn{2}{c}{$2.80^{+0.20}_{-0.10}$} \\\\

{\sc nthComp} & $\Gamma$ & $1.66\pm0.03 $ & $2.33 \pm 0.03
                                             $ & $1.70 \pm 0.02$ &
                                                                   $2.123^{+0.026}_{-0.013} $& 
$1.71 \pm 0.05$ & $2.155 \pm 0.005 $ & $1.69 \pm 0.02$  & $2.152\pm0.015$\\
 & kT$_{\rm e}$ (keV) & $1.92 \pm 0.04$ &  $4.33 \pm 0.10$ & $2.04 \pm
                                                             0.04$ &
                                                                     $3.56 \pm 0.05$& 
$2.022 \pm 0.012$ & $3.69 \pm 0.02$&$2.00 \pm 0.03$& $3.66\pm0.04  $ \\
 & kT$_{\rm bb}$ (keV) & $0.55 \pm 0.03$ &  $0.721\pm0.015$
                               & $0.44 \pm 0.03$ & $0.578  ^{+0.027}_{-0.014} $& 
$0.446 ^{+0.020}_{-0.013} $ & $0.577^{+0.012}_{-0.006} $&$0.434^{+0.024}_{-0.015}$&  $0.573^{+0.015}_{-0.008}$\\
 & Norm & \multicolumn{2}{c}{$0.063 \pm 0.004$}&
                                                         \multicolumn{2}{c}{$0.20\pm0.03$}
                               & \multicolumn{2}{c}{$0.148 \pm 0.012$} & \multicolumn{2}{c}{$0.145 \pm 0.012$}\\\\

 & $\chi^2/dof$ & \multicolumn{2}{c}{4121/2801} & \multicolumn{2}{c}{3331/2797} & \multicolumn{2}{c}{3252/2794} & \multicolumn{2}{c}{ 3228/2794} \\
\hline
\hline
\end{tabular}
\caption{Best-fit parameters of the high-flux spectrum. The spectra
  associated with the 2001 observation are the RGS12 and EPN spectra while the
  spectra associated with   the 2016 observation are the \swift and
  \nustar ones.}
\label{tab:high_flux}
\end{table*}  

We adopted the cosmic abundances and the cross sections
derived by \cite{Wilms_00} and \cite{Verner_96}, respectively. With the
aim to account for the interstellar absorption, we adopted the
Tü\"ubingen-Boulder model ({\sc TBabs} in XSPEC).  The uncertainties are
reported at 90\% confidence level (c.l.).

\subsection{High-flux spectrum (the soft state)}
\label{sec:spectral analysis high}

\begin{figure}[ht]
\includegraphics[width=8.5cm]{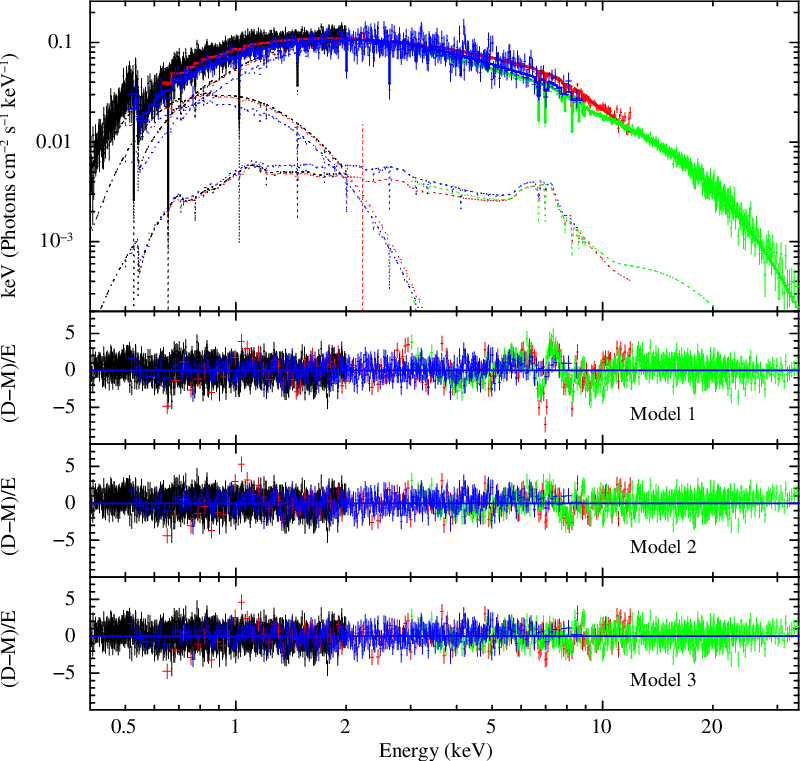} 
\caption{High-flux spectrum and residuals corresponding to the models
  discussed in the text. The black, red, green, and blue data are
  associated with the RGS12, EPN, \nustar, and \swift spectra,
  respectively.  From top to bottom: $\rm{E*f(E)}$ is the unfolded
  spectrum associated with {\it \textup{Model 3}}, and residuals in units of $\sigma$
  ([data-model]/error) associated with  {\it \textup{Model 1, Model 2 and Model 3}},
  respectively.}
\label{fig:res_soft}
    \end{figure}

We adopted a model that takes into account an emission from
the accretion disk \citep[{\sc diskbb} in XSPEC, see][]{Mitsuda_84}
plus a thermally Comptonized component \citep[{\sc nthComp} in XSPEC,
see][]{Zic_99} to fit the continuum emission.
Because the RGS12 and EPN events were collected at
different times than the \swift  and \nustar  events (i.e.,
2001 and 2016, respectively),
we left the seed
photon temperature $kT_{\rm bb}$, the electron temperature
$kT_{\rm e}$, and the asymptotic power-law photon index
$\Gamma$ free to vary independently. Moreover, we fixed the parameter {\sc inp\_type} of
{\sc Nthcomp} to zero, assuming that the seed photons have a blackbody
spectrum (i.e., the seed photons originate from the NS
surface).  Because large residuals were evident   at 0.54 keV in
  the RGS12 spectrum,
we added an absorption edge only to the RGS12 spectrum. This feature 
is interpreted as a calibration issue associated with the K edge of
neutral oxygen in the RGS spectra \citep{devri_03}.
 We modeled the Au instrumental edge ($\sim 2.2$
keV) in the EPN spectrum adding a Gaussian component, as was done by \cite{Pintore_16}
and \cite{Sanna_17}. We fixed the
energy of the line  at
2.22 keV and imposed a null width. The Gaussian normalization was left free
to vary for the EPN spectrum but was fixed to zero for the other spectra. 
Then the initial adopted model was
$$
 {\rm { Model\; 1 = Edge*TBabs*(diskbb+nthComp)}}.
 $$
 We found a $\chi^2(d.o.f)$ of 4121(2801). We show the best-fit
 parameters of Model 1 in the third column of
 Tab. \ref{tab:high_flux}. The residuals are shown in the second
 panel from the top in  Fig. \ref{fig:res_soft}.  We observed that large residuals were
 present between 6 and 10 keV. These can be interpreted as absorption
 lines and edges associated with the presence of \ion{Fe}{xxv} and
 \ion{Fe}{xxvi} ions.  Furthermore, the RGS12 spectrum showed two
 absorption lines at 0.65 and 1.02 keV that might be associated with
 \ion{O}{viii} and \ion{Ne}{x} ions.

 To fit these absorption features, we added a
 multiplicative component, {\sc zxipcf}\footnote{https://heasarc.gsfc.nasa.gov/xanadu/xspec/models/zxipcf.html}, to Model 1 that takes into account a
 partial covering of ionized absorbing material. This component
 reproduces the absorption from photoionized matter that is illuminated by a
 power-law source with spectral index $\Gamma=2.2$ and assumes that
 the photoionized absorber has a microturbulent velocity of 200 km
 s$^{-1}$. The component {\sc zxipcf} was initially developed to
 study absorption features in the Fe-K region in AGN
 \citep[see][]{Reeves_08} and to study the complex spectra of narrow-line Seyfert 1 galaxies \citep[see][]{Miller_07}.  Recently, this
 component was adopted to fit the absorbing features in the Fe-K
 region that is associated with highly ionized matter
 surrounding the eclipsing NS-LMXB AX J1745.6-2901 \citep{Ponti_15}
 and the dipping source XB 1916-053 \citep{Gambino_19}.
 The parameters of this component are N$_{\rm H}$, log${(\xi)}_{\rm IA}$, $f _{\rm IA}$,
 and $z$. N$_{\rm H}$ describes the equivalent hydrogen column density
 associated with the ionized absorber, and log${(\xi)}_{\rm IA}$ describes the
 ionization degree of the absorbing material. It is defined as
 ${\xi}_{\rm IA} = L_x/(n_Hr^2)$, where $L_x$ is the X-ray luminosity
 incident on the absorbing material, $r$ is the distance of the absorber
 from the X-ray source, and $n_H$ is the hydrogen atom density of the
 absorber. $f _{\rm IA}$ indicates the fraction of the source covered by the
 absorber, and $z$ gives the redshift of the source and was fixed   to zero.

 To take into account that the ionized absorber can scatter the
 radiation out of  the line of sight via Thomson or Compton scattering, we
 added the multiplicative component {\sc cabs} to the model. The only
 parameter of the component {\sc cabs} is N$_{\rm H}$, which describes
 the equivalent hydrogen column density associated with the scattering
 cloud. We tied the value of this parameter to that of the equivalent
 hydrogen column density associated with the absorbing ionized cloud.

In addition,  in order to take into account that the absorbing material might cover
 only a fraction $f$ of the source, we multiplied the component {\sc
   cabs} by the component {\sc partcov}. The latter is a convolution model
 that allows converting an absorption component into a partially
 covering absorption component. The parameter of {\sc partcov} is $f$, that
 is, the fraction of the source covered by the absorbing material; we
 tied its value to that of the parameter $f$ of the component {\sc
   zxipcf}. Furthermore, we left the ionization
 parameter of the component {\sc zxipcf} for the spectra taken in 2001
 and for those taken in 2016 free to vary independently.  In summary,  {\it \textup{Model 2}} can be
 described as
\begin{equation*}
\begin{split}
 {\rm { Model\; 2 = Edge*TBabs*(partcov*cabs)*}} \\
 {\rm { zxipcf*(diskbb+nthComp)}}.
 \end{split}
\end{equation*}

By fitting the spectrum, we obtained a $\chi^2(d.o.f)$ of 3331(2797)
and decreases   of  790.  The
addition of the {\sc zxipcf} component is statistically significant,
as witnessed by the extremely low outcome of the F-test probability of
chance improvement with respect to {\it \textup{Model 1}}, that is,$1.5 \times 10^{-127}$. The
residuals associated with the absorption lines at 0.65 and 1.02 keV
disappeared, while those in the 6-10 keV energy range were largely
reduced.  We show the best-fit values of the parameters in the fourth
column of Tab. \ref{tab:high_flux}, and the residuals are shown in the
third panel from the top in Fig. \ref{fig:res_soft}.  The equivalent
hydrogen column associated with the ionized absorbing material is
$(1.2 \pm 0.2) \times 10^{24}$ cm$^{-2}$, which is a factor  400 larger
than the equivalent hydrogen column associated with the neutral
interstellar matter of $(2.82 ^{+0.08}_{-0.13}) \times 10^{21}$ cm$^{-2}$. The covering
fraction of the absorber is $0.88 \pm 0.07$, suggesting that
a large part of the source is occulted by the absorber. The
ionization parameter log$(\xi)_{\rm IA}$  is $4.45^{+0.14}_{-0.09}$ and
$4.24 \pm 0.05$ erg cm s$^{-1}$ for the 2001 and 2016 observation,
respectively. 

Residuals were still  present  between 6 and 9 keV. These might be
associated with a reflection component from the
accretion disk. \cite{Sidoli_01} detected  a broad
emission line in the Fe-K region that might be associated with 
a fluorescence iron line originating from disk reflection.
In order to verify the presence of a broad emission line, we added a
Gaussian component to {\it \textup{Model 2}} and again fit the broadband
spectrum. We found a $\chi^2(d.o.f)$ of  3252(2794) and the F-test probability of chance
improvement was $2 \times 10^{-14}$. This shows that the addition of this
component is statistically significant.  The energy of the broad emission line is $6.60 \pm 0.10$
keV, the width is $0.69 \pm 0.08$ keV, and the normalization is
$(1.1 \pm 0.1) \times 10^{-3}$ photons cm$^{-2}$ s$^{-1}$.  These
results are compatible within a 90\% confidence level with those reported
by \cite{Sidoli_01}. 

In the following  we investigate a scenario in which
the broad emission line is produced by reflection from the inner
region of the accretion disk, where the relativistic effects smear the
reflection component.
We convolved the Comptonized component {\sc
  nthComp} with the reflection component {\sc rxfconv} \citep[see][for
details]{Done_06,Kole_2011}.  We imposed in our
model that the incident luminosity comes from the Comptonized component.
The component {\sc rxfconv} has five
parameters: the first is the redshift $z,$ which we kept fixed to
zero, the second is the iron abundance Fe$_{\rm abundance}$ , which
we kept fixed to the solar abundance, the third is the cosine of
the inclination angle $\theta$ of \source, which we assumed to be
72$^{\circ}$ \citep{Iaria_18}, the fourth is the
ionization parameter of the reflecting surface of the accretion disk
log$(\xi)$, which we left free to vary, and, finally, the fifth
parameter is the relative reflection normalization of the component
{\sc rxfconv}, measured in units of solid angle $\Omega/2\pi$ subtended
by the reflector as seen from the corona (rel$_{\rm refl}$, hereafter).  

However, the reflection region might be close to the
NS, therefore  we added the component {\sc rdblur} to take into account
the smearing associated with general and special relativistic
effects.  {\sc rdblur} has four  parameters: the inclination
angle of the source, which we fixed at 72$^{\circ}$ (see above), the
inner and outer radius, R$_{\rm in}$ and R$_{\rm out}$, of the
reflection region in units of gravitational radii ($GM/c^2$, where $M$
is the mass of the compact object), and, finally, the power-law
dependence of emissivity, Betor10.  

In summary,   {\it \textup{Model 3}} is composed of
\begin{equation*} 
\begin{split}
 {\rm { Model\; 3 = Edge*TBabs*(partcov*cabs)* zxipcf*}} \\
 {\rm {(diskbb+rdblur*rfxconv*nthComp)}}.
 \end{split}
\end{equation*}

    Initially, we left R$_{\rm in}$, R$_{\rm out}$  , and
    Betor10 free to vary.  The value of $f_{\rm IA}$ shifts toward 1, suggesting that the
    whole X-ray source is shielded by the ionized absorber. Hence we
    fixed the value of $f_{\rm IA}$ at 1. Furthermore, because the value of $\chi^2$ is
    less sensitive to the changes of R$_{\rm out}$, we fixed the value
    of R$_{\rm out}$ at 2800 gravitational radii, which is the best-fit
    value of R$_{\rm out}$.

    A $\chi^2(d.o.f)$ of  3228(2794) was obtained after    the reflection component was added, whose statistical significance was
    confirmed by the F-test probability of chance improvement with
    respect to {\it \textup{Model 2}} equal to $7 \times 10^{-19}$. We also
    found that the self-consistent reflection model further  improves the fit, resulting in a decrease of $\chi^2$ 
       of  24 with respect to {\it
      \textup{Model 2 plus broad line}}.  We show the best-fit values in the
    sixth column of Tab. \ref{tab:high_flux}. The unfolded
    $\rm {E*f(E)}$ spectrum and the corresponding residuals are shown
    in the top and bottom panels in Fig. \ref{fig:res_soft},
    respectively.

    We find that the equivalent hydrogen column density associated
    with the neutral interstellar matter is
    $(2.75 \pm 0.10) \times 10^{21}$ cm$^{-2}$. The ionized absorbing
    matter has an equivalent hydrogen column density of
    $(6.0^{+0.9}_{-1.4}) \times 10^{23}$ cm$^{-2}$ , and its ionization
    parameter log$({\xi})_{\rm IA}$ was found equal to $4.37 \pm 0.04$ and
    $4.14 \pm 0.05$ for Obs. 2001 and Obs. 2016, respectively.  The
    inner temperature of the multicolored disk emission is
    $0.30 \pm 0.02$ keV and the corresponding inner radius is
    R$_{\rm disk}\sqrt{\cos \theta} = 39\pm4$ km, assuming a
    distance  of 10 kpc. The fit revealed that the
    power-law index and the electron and seed photon temperatures are all
    higher in Obs. 2016 than in Obs. 2001 (see
    Tab. \ref{tab:high_flux}).

    The reflection component has a relative normalization of
    $0.22^{+0.12}_{-0.05} $, the inner radius of the reflecting region is
    $60^{+60}_{-30}$ gravitational radii, and the power-law dependence
    of emissivity is lower than -1.4. Finally, the 0.4-35 keV absorbed flux is $1.2 \times 10^{-9}$ erg
cm$^{-2}$ s$^{-1}$, the extrapolated unabsorbed  (excluding both the neutral
and ionized absorber)  flux in the 0.1-100
keV energy range is $2.2 \times 10^{-9}$ erg cm$^{-2}$ s$^{-1}$ , and
the extrapolated unabsorbed flux in the 0.1-100 keV energy range of
the Comptonized component is $1.9 \times 10^{-9}$ erg cm$^{-2}$
s$^{-1}$. Assuming a distance   of 10 kpc, the 0.1-100 keV
extrapolated luminosity of the Comptonized component is
$2.2 \times 10^{37}$ erg s$^{-1}$.

{\it \textup{Model 3}} assumes that the broadening of the emission line observed
in the Fe-K region of the spectrum is caused by both Compton
scattering (included in the {\sc rfxconv} component) and relativistic
smearing (included in the {\sc rdblur} component).  We checked whether the
Compton scattering might by  itself explain the  width of the emission
line by removing the component {\sc rdblur} from {\it \textup{Model 3}}. 
This seems not to be the case, as pointed out by the   slightly higher
$\chi^2(d.o.f)$ of  3254(2796) with respect to  the
$\chi^2(d.o.f)$   value of 3228(2794)  found for  {\it \textup{Model 3}} and 
the F-test probability of chance
improvement of $1.4 \times 10^{-5}$ , corresponding to a statistical 
improvement larger than 4 $\sigma$.

 \begin{figure}[ht]
\includegraphics[width=8.5cm]{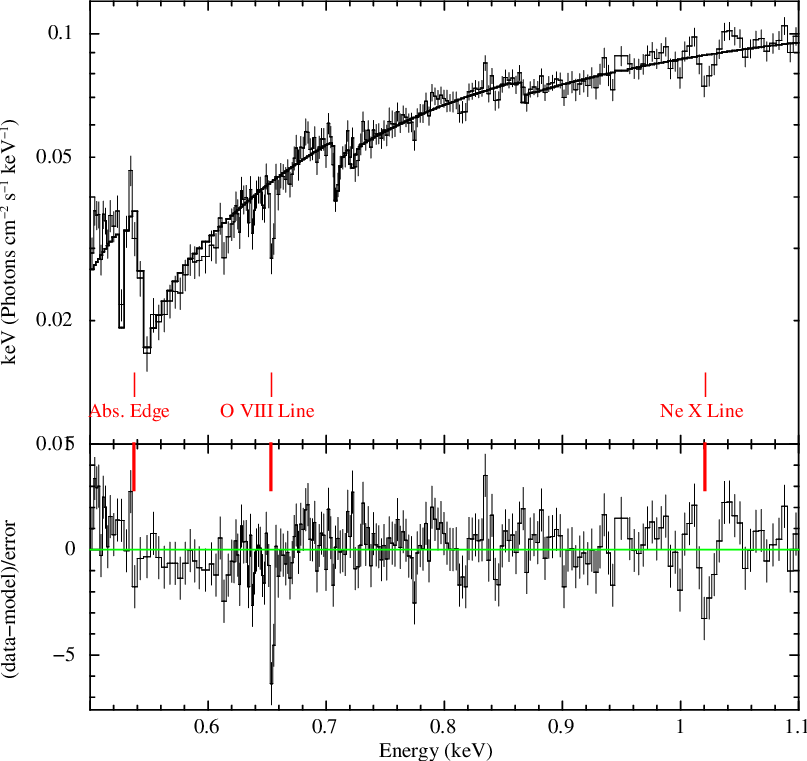} 
\caption{High-flux RGS12 spectrum and residuals corresponding to the models
composed of a power-law component absorbed by neutral interstellar
matter.  We marked the energies at 0.54, 0.64, and 1.022 keV
that correspond to the absorption edge associated with neutral oxygen and to
the absorption lines associated with  \ion{O}{viii}  and \ion{Ne}{x},
respectively.}
\label{fig:rgs_spec}
    \end{figure}

The RGS12 and EPN spectrum showed clear signatures of absorption lines at
0.65 keV, 1 keV, and in the Fe-K region. 
  We show in Fig. \ref{fig:rgs_spec} the RGS12 spectrum in the 0.5-1.1 keV energy range 
   and the corresponding residuals obtained by  fitting the data with a 
  power-law component with a photon index of $1.25 \pm 0.04$, absorbed by
  neutral interstellar matter with an equivalent hydrogen column
  density of $(0.302 \pm 0.008) \times 10^{22}$ cm$^{-2}$.
We considered the RGS12 energy
range between 0.62 and 0.70 keV and fit the continuum emission
by adopting a power-law component. Then we added a Gaussian line with
negative normalization to fit the absorption line at 0.65 keV, and we
fixed the value of the power-law index to its best-fit value to
estimate the uncertainties for the absorption line parameters. The best-fit
parameters of the Gaussian absorption lines are shown in
Tab. \ref{tab:line}. Using the RGS12 energy range between 0.98 and
1.04 keV and adopting the same approach described above, we found the
best-fit parameters of the Gaussian absorption line associated with
the \ion{Ne}{x} Ly$_{\alpha}$ transition.  We selected the EPN energy range
between 6.4 and 7.5 keV and found the best-fit parameters of the
Gaussian absorption lines to be associated with the resonant transition of
\ion{Fe}{xxv}  line and to the \ion{Fe}{xxvi} Ly$_{\alpha}$
line. Finally, selecting the 7.4-8.5 keV energy range, we estimated
the best-fit values of the Gaussian absorption lines associated with
the Ly$_{\beta}$ transitions of the \ion{Fe}{xxv} and
\ion{Fe}{xxvi}  lines. All these results are shown in
Tab. \ref{tab:line}.
\begin{table*} [!htbp]
\centering
\footnotesize
\begin{tabular}{lccccccc}
\hline
\hline
Ion and  & Measured & Theoretical & EW & Intensity & \multicolumn{2}{c}{Line Width} \\
Transition & Energy (keV) & Energy (keV) & (eV) & ($10^{-4}$ ph cm$^{-2}$ s$^{-1}$) & (eV) & (km s$^{-1}$)\\
\hline
\ion{O}{viii} 1s-2p & $0.6540 \pm 0.0003$& 0.6536 & $-1.8^{+0.3}_{-0.4}$& $-1.2 \pm 0.2$& $<1.2$&$<550$\\
\ion{Ne}{x} 1s-2p &$1.0216^{+0.0019}_{-0.0011}$ & 1.0218 & $-3.6 \pm 1.0 $ & $-3.4^{+0.7}_{-0.5} $& $2.8^{+1.3}_{-0.8}$ &$820^{+380}_{-230}$\\
\ion{Fe}{xxv} 1s$^2$-1s2p & $6.705^{+0.0160}_{-0.0013}$ & 6.700 & $-34^{+6}_{-4} $& $-2.5 \pm 0.3$& $<42$&$<1900$\\
\ion{Fe}{xxvi} 1s-2p &  $6.992^{+0.015}_{-0.004}$& 6.966 & $-46^{+5}_{-3}$& $-3.1 \pm 0.3$& $<31$&$<1300$\\
\ion{Fe}{xxv} 1s$^2$-1s3p &  $7.91^{+0.06}_{-0.04}$& 7.88 & $-33 \pm 11$ & $-1.6^{+0.5}_{-0.7}$&  $<190$& $<7200$\\
\ion{Fe}{xxvi} 1s-3p & $8.26 \pm 0.03$ & 8.25 & $-45 \pm 10$ & $ -1.9 \pm 0.5$ & $<130$& $<4700$ \\
 \hline
\hline
\end{tabular}
\caption{Best-fit parameters of the absorption lines in the high-flux
  spectrum. Uncertainties are given at 68 \% c.l., upper
      limits at 90\% c. l.. The best-fit values of the energies are
      compatible with the rest-frame values at 90\% c. l.}
\label{tab:line}
\end{table*}

 \begin{table*} [!htbp]
\caption{Best-fit parameters of the low-flux spectrum}
\label{tab:low_flux}
\centering
\footnotesize
\begin{tabular}{llcccccc}
\hline
\hline
Model & Component &\multicolumn{2}{c}{Model 1} &\multicolumn{2}{c}{Model 2} &\multicolumn{2}{c}{Model 3} \\
& &  \xmm &  \nustar &  \xmm & \nustar &  \xmm &  \nustar\\ 
\hline
{\sc zxipcf} & N$_{\rm H}$  (10$^{22}$ atoms cm$^{-2}$) & \multicolumn{2}{c}{-}& \multicolumn{2}{c}{$13 \pm 2$} & \multicolumn{2}{c}{$14.1 \pm 1.2 $} \\
 &log$(\xi)_{\rm IA}$ & \multicolumn{2}{c}{-} & $1.09^{+0.12}_{-0.28} $ &  $2.9 \pm 0.2$& $1.98 \pm 0.05$ &  $3.1 \pm 0.3$\\
 & $f_{\rm IA}$ & \multicolumn{2}{c}{-} & \multicolumn{2}{c}{$0.21 \pm 0.02$} & \multicolumn{2}{c}{ $0.25 \pm 0.02$} \\\\

{\sc TBabs} & N$_{\rm H_{ISM}}$ (10$^{22}$ atoms cm$^{-2}$) & \multicolumn{2}{c}{$0.231 \pm 0.005$} & \multicolumn{2}{c}{$0.237^{+0.007}_{-0.013}$} & \multicolumn{2}{c}{$0.329 ^{+0.004}_{-0.008}$ }\\\\

{\sc partcov} & $f$ & \multicolumn{2}{c}{-} & \multicolumn{2}{c}{$0.21
                                              \pm 0.02$} &
                                                           \multicolumn{2}{c}{
                                                           $0.25 \pm
                                                           0.02$} \\ 
{\sc cabs} & N$_{\rm H}$  (10$^{22}$ atoms cm$^{-2}$) & \multicolumn{2}{c}{-}& \multicolumn{2}{c}{$13 \pm 2$} & \multicolumn{2}{c}{$14.1 \pm 1.2 $} \\\\

{\sc rdblur} & Betor10 & \multicolumn{2}{c}{-}& \multicolumn{2}{c}{-}& \multicolumn{2}{c}{ $-3.0 \pm 0.3$} \\
 & R$_{\rm in}$ $\rm (GM/c^2)$ & \multicolumn{2}{c}{-}& \multicolumn{2}{c}{-} & \multicolumn{2}{c}{ $<7$ } \\
 & R$_{\rm out}$ $\rm (GM/c^2)$ & \multicolumn{2}{c}{-}& \multicolumn{2}{c}{-} & \multicolumn{2}{c}{$290$ (frozen)} \\
 & $\theta$ (deg) & \multicolumn{2}{c}{-}& \multicolumn{2}{c}{-}   & \multicolumn{2}{c}{ 72 (frozen)} \\\\

{\sc rfxconv} & rel$_{\rm refl}$& \multicolumn{2}{c}{-}& \multicolumn{2}{c}{-} & \multicolumn{2}{c}{$0.48 \pm 0.06$} \\
 & Fe$_{\rm abund}$ & \multicolumn{2}{c}{-}& \multicolumn{2}{c}{-} & \multicolumn{2}{c}{1 (frozen)} \\
 & $\cos \theta$ & \multicolumn{2}{c}{-}& \multicolumn{2}{c}{-} & \multicolumn{2}{c}{0.309 (frozen)}\\
 & log($\xi$) & \multicolumn{2}{c}{-}& \multicolumn{2}{c}{-} & \multicolumn{2}{c}{$1.99^{+0.05}_{-0.10}$ } \\\\

{\sc nthComp} & $\Gamma$ & \multicolumn{2}{c}{$1.842\pm0.004$} & \multicolumn{2}{c}{$1.911 \pm 0.011$}& \multicolumn{2}{c}{$1.989 \pm 0.015$} \\
 & kT$_{\rm e}$ (keV) & \multicolumn{2}{c}{$37^{+35}_{-10}$}& \multicolumn{2}{c}{$>180$} & \multicolumn{2}{c}{$>130$} \\
 & kT$_{\rm bb}$ (keV) & \multicolumn{2}{c}{$0.113 \pm 0.007$} & \multicolumn{2}{c}{$0.120 \pm 0.009$} & \multicolumn{2}{c}{$<0.06$}\\
 & norm & \multicolumn{2}{c}{$0.0242 \pm 0.0003$} & \multicolumn{2}{c}{$0.0318  \pm 0.0006$} & \multicolumn{2}{c}{$0.0343 \pm 0.0011$} \\\\

 & $\chi^2/dof$ & \multicolumn{2}{c}{2967/2278} & \multicolumn{2}{c}{2556/2274} & \multicolumn{2}{c}{2359/2270} \\
\hline
\hline
\end{tabular}
\end{table*}  

\subsection{Low-flux spectrum (the hard state)}
\begin{figure}[ht]
\includegraphics[width=8.5cm]{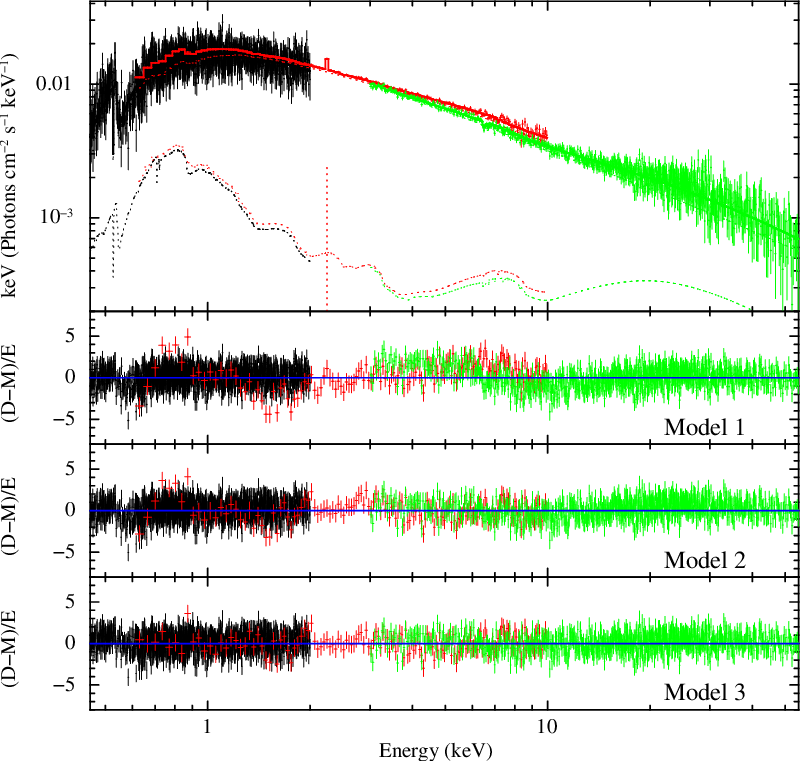} 
\caption{Low-flux spectrum and residuals corresponding to the models
  discussed in the text. The colors are defined as in
  Fig. \ref{fig:res_soft}.}
\label{fig:res_hard}
    \end{figure}
    We used the same    {\it \textup{Model 1}} as we adopted for the
    high-flux spectrum, but the {\sc diskbb} component is not
    required here. To fit the low-flux spectrum, we also added a Gaussian
    emission line at 2.22 keV to fit the residuals of the EPN
    spectrum between 1.8 and 2.4 keV.  The first model in this case is
\begin{equation*}
{\rm { Model\; 1 = TBabs*nthComp.}}
 \end{equation*}
 We obtained a $\chi^2(d.o.f)$ value of   2967(2278).  We show the
 best-fit parameters of {\it \textup{Model 1}} in the third column of Tab. \ref{tab:low_flux},
 and the residuals are shown in the second panel from the top in
 Fig. \ref{fig:res_hard}.  The residuals show
 absorption features between 6 and 9 keV. For this reason we added the
 {\sc zxipcf} component. The new model is
\begin{equation*}
\begin{split}
 {\rm { Model\; 2 = TBabs*(partcov*cabs)*}} \\
 {\rm { zxipcf*nthComp}}.
 \end{split}
\end{equation*}
We left the ionization parameter associated with
the \xmm and \nustar spectrum free to vary independently.
Fitting the spectra with {\it \textup{Model 2,}} we obtained a $\chi^2(d.o.f)$ value
of 2556(2274). The addition of the {\sc zxipcf} component is
statistically significant (the F-test probability of chance
improvement is $4 \times 10^{-72}$). We show the best-fit parameters in
the fourth column of Tab. \ref{tab:low_flux}. However, we observe that large residuals are still present above 10 keV in the \nustar spectrum (see the third panel from the top of Fig. \ref{fig:res_hard}).
These might be associated with a Compton
hump caused by reflection from the accretion disk. Similarly to the case of  
 the high-flux spectrum, we added a reflection component
smeared by relativistic effects. The adopted model was
\begin{equation*}
\begin{split}
 {\rm { Model\; 3 = TBabs*(partcov*cabs)*}} \\
 {\rm { zxipcf*rdblur*rfxconv*nthComp}}.
 \end{split}
\end{equation*}
The value of $\chi^2$ is less sensitive to change in R$_{\rm out}$,
hence we
fixed the value of R$_{\rm out}$ to the best-fit value of 290
gravitational radii. We obtained a $\chi^2(d.o.f)$ value of
 2359(2270). The addition of the reflection component is statistically
significant, with an obtained  F-test probability of chance improvement of
$3 \times 10^{-38}$. We show the best-fit values in the fifth column
of Tab. \ref{tab:low_flux}, and  the unfolded $\rm {E*f(E)}$ spectrum with
the model components and the corresponding residuals are shown in the
top and bottom panels in Fig. \ref{fig:res_hard}, respectively.

We  refit   the spectrum by fixing the value of R$_{\rm out}$ to the value
obtained for the high-flux spectrum (2800 gravitational radii); the
best-fit values of the other parameters did not change. However, the
$\chi^2$ value is higher and $\Delta \chi^2$ is 4, therefore we
discuss our results with R$_{\rm out}$ fixed at 290 gravitational radii.

We obtained that the equivalent hydrogen column associated with the
neutral interstellar matter is
N$_{\rm H_{ISM}} = (3.29^{+0.04}_{-0.08}) \times 10^{21}$ cm$^{-2}$. The ionized
absorbing matter around the system has an equivalent hydrogen column
of N$_{\rm H} = (1.41 \pm 0.12) \times 10^{23}$ cm$^{-2}$ and it covers
$(25 \pm 2)$\% of the emitting source. The value of log$(\xi)_{\rm IA}$ is 
 $1.98 \pm 0.05$ and $3.1 \pm 0.3$ for the \xmm
and \nustar observation, respectively.  The Comptonized component is
hard,  with an  electron temperature kT$_{\rm e}$ higher
than 130 keV and   a seed-photon temperature kT$_{\rm bb}$ lower 
than 0.06 keV; the photon index is  $\Gamma = 1.989 \pm 0.015$.

The best-fit values for log$(\xi)$ and rel$_{\rm refl}$
($1.99^{+0.05}_{-0.10}$ and $0.48 \pm 0.06$, respectively) suggest
that the reflecting region probably lies close the NS and that
the Comptonized corona may have a slab geometry.  The power-law
dependency of emissivity is Betor10 $ =-3.0 \pm 0.3,$ and the inner
radius of the reflecting region is smaller than seven gravitational radii.

Finally, the 0.45-55 keV absorbed flux is $2 \times 10^{-10}$ erg
cm$^{-2}$ s$^{-1}$ and the extrapolated unabsorbed  (excluding both the neutral
and ionized absorber) flux in the 0.1-100
keV energy range is $4.4 \times 10^{-10}$ erg cm$^{-2}$ s$^{-1}$. The
extrapolated unabsorbed flux in the 0.1-100 keV energy range of the
Comptonized component is $4 \times 10^{-10}$ erg cm$^{-2}$ s$^{-1}$
and the corresponding luminosity  is $4 \times 10^{36}$ erg s$^{-1}$ for a distance to the source of 10kpc.

\section{Discussion}
\label{sec:discussion}

We have analyzed the high-flux and low-flux spectra of \source collected   during
the 1999 and 2015 outbursts.  We find that the high-flux spectrum
shows  a soft Comptonized component with a
value of the electron temperature lower  than 4 keV. Furthermore, a
multicolor  disk blackbody component is present at low energies with an inner
temperature of 0.30 keV and an inner radius of the disk of
R$_{\rm in}\sqrt{\cos \theta} = 39 \pm 4$ km (assuming a distance
to the source of 10 kpc).  The low-flux spectrum has a
hard Comptonized component with a value of the electron temperature
higher than 130 keV, while the addition of a multicolor  disk component is
not statistically significant.

According to the fits, in both observations the source is (at
least partially) covered by ionized absorbing matter. 
In particular, the covering by this absorbing material is complete in the
soft state, where the equivalent hydrogen column density is  found
equal to  $(60^{+9}_{-14}) \times 10^{22}$ cm$^{-2}$,
 and only partial in the hard state, with a covering fraction of
 25\%. 
The ionization parameter log$({\xi})_{\rm IA}$ of the absorbing material is higher
than 4 in the high-flux state and lower than 3.1 in the low-flux state.

    The spectra both show a reflection component from the accretion
    disk. We find that when we take the relativistic smearing into account by adding the {\sc rdblur}
    component to the reflection component, we improve the fit with a
    statistical significance higher than 4 $\sigma$.  This suggests
    that the broad width of the emission line observed in the Fe-K
    region cannot be explained invoking only the Compton broadening,
    but it is necessary to also take the relativistic
    smearing into account.

    The reflecting skin above the disk has an ionization parameter
    log$(\xi)$ of $2.80^{+0.20}_{-0.10}$ and $1.99^{+0.05}_{-0.10}$ in
    the soft and hard state, respectively.  The reflecting region is
    between $39^{+35}_{-15}$ and 2800 gravitational radii in the high
    state, and it is closer to the NS surface in the low state,
    for which its boundaries are smaller than 7 and 290 gravitational
    radii, respectively.  A similar behavior of the reflection region 
    was observed by \cite{Mazzola_19} in soft and high state of the NS-LMXB 4U 1702-429 and by \cite{Disalvo_15} in 4U 1705-44 \citep[see also][]{Egron_13}.

    Assuming an NS mass of 1.54 M$_{\odot}$
    \citep[see][]{Ozel_16}, the reflecting region in the high state is
    between $140^{+140}_{-70}$ km and $6.4 \times 10^3$ km.  The inner
    radius of the accretion disk is R$_{\rm disk} = 70\pm7$ km,
    under the assumption that the inclination angle of \source is
    72$^{\circ}$. However, because the inferred
    luminosity (see Sect. \ref{sec:spectral analysis high}) is 10\% of
    the Eddington luminosity, the effective inner radius of the
    accretion disk r$_{\rm disk}$ is given by the relation
    r$_{\rm disk} = s^2 {\rm R_{\rm disk} }$ where $s=1.7$
    \citep[see][]{Shimura_1995}.  We obtain  
    r$_{\rm disk} = 200\pm20$ km, which is compatible to the
    boundaries of the reflecting region estimated above.

We estimated the optical depth of the Comptonized cloud for the high and
soft state using the relation between the power-law photon index
$\Gamma$ and the electron temperature kT$_{\rm e}$ obtained by
\cite{Zidi_96}. As expected, we find that the Comptonizing cloud is optically thick in the soft state ($\tau$ is  $15.7 \pm 0.5$  and $8.0 \pm 0.2$  for the spectra taken in  2001 and 2016, respectively), and it is optically thin in the hard state ($\tau \lesssim  0.7$).

We  determined the average electron number density n$_{\rm e}$ of the
Comptonized cloud using the relation $\tau = {\rm \sigma_T n_e d}$,
where ${\rm \sigma_T }$ is the Thomson cross-section and ${\rm d}$ is
the geometrical dimension of the cloud. In the case of the soft state
we assumed that ${\rm d}$ is equal to r$_{\rm disk}$, which is
plausible because the disk emission from the innermost region could be occulted by the
optically thick   Comptonized cloud surrounding the compact object.
We found n$_{\rm e} = (9 \pm 2) \times 10^{17}$ cm$^{-3}$ and
n$_{\rm e} = (5 \pm 1) \times 10^{17}$ cm$^{-3}$  for the soft state
during the observation taken in 2001 and 2016, respectively.
An upper limit on   the average electron
number density n$_{\rm e}$ of the Comptonized cloud in the hard state 
can be inferred assuming that the size of the Comptonized cloud has
to be smaller than the outer radius of the reflecting skin (290
gravitational radii, corresponding  to 670 km for an NS mass
of 1.54 M$_{\odot}$), assuming that the Comptonized cloud does
not fully cover the reflecting region.  Knowing that the optical depth
associated with  the Comptonized cloud is lower than 0.7, we obtain 
n$_{\rm e} < 2 \times 10^{16}$ cm$^{-3}$ at a distance  of 670 km from
the NS.

On the other hand, the electron number density n$_{\rm e}$ associated
with the reflecting skin can be inferred taking into account that
log$(\xi) \simeq 2.8$, because $\xi=L_x/(n_er^2)$, where $L_x$ is the
incident luminosity and $r$ is the distance between the X-ray source
and the reflecting region. The incident luminosity $L_x$ 
associated with the Comptonized component is
$2.4 \times 10^{37}$ erg s$^{-1}$ in the high state, and $r$ is roughly
between $140^{+140}_{-70}$ km and $6.4 \times 10^3$ km. By substituting
these values, we infer that the electron number density of the
reflecting skin is between $\sim 2 \times 10^{20}$ and
$\sim 9 \times 10^{16}$ cm$^{-3}$ going from R$_{\rm in}$ to
R$_{\rm out}$; at the distance of r$_{\rm disk}$ , we find that
n$_{\rm e} \simeq 1.0 \times 10^{20}$ cm$^{-3}$.  The electron number
density n$_{\rm e}$ associated with the reflecting skin in the
low state is between $\sim 4 \times 10^{22}$ and
$\sim 1 \times 10^{19}$ cm$^{-3}$ going from the NS surface
 to R$_{\rm out} =670$ km.

 The best-fit value of the rel$_{refl}$ parameter, that is, the  normalization of the {\sc rxfconv} component, is
$0.22^{+0.12}_{-0.05}$ and $ 0.48 \pm 0.06$ for the high-flux spectrum (optically
thick corona) and low-flux spectrum (optically thin corona),
respectively.  These values describe   a scenario well in which a 
spherical corona is present in the inner part of the accretion disk
\citep[see Fig. 5 in][]{Dove_97}. The reflecting radius
  is a factor 1.2 larger than the coronal radius in the high-flux state, and the two radii are compatible with each other in the low-flux state.  


We detect an ionized absorber in both the high and
low state.  Assuming roughly that $n_{\rm e} =n_{\rm H}$, we expect
that the absorbing ionized cloud is optically thin. Solving 
$\tau=\sigma_T n_{\rm e} d =\sigma_T n_{\rm H} d = \sigma_T N_{\rm H}
$, we obtained $\tau \simeq 0.4$ and $\tau \simeq 0.09$ for the
soft and hard state, respectively.  To describe the ionized absorber, we
 adopted the model  {\sc zxipcf}  and fit the data assuming a unique
value of the ionization parameter, unlike  \cite{Sidoli_01}, who  suggested that
the ionized matter around \source should have a gradient of ionization
to explain the simultaneous presence of absorption lines associated with
heavy ions such as \ion{Fe}{xxv} and \ion{Fe}{xxvi} and light ions such as
\ion{O}{viii} and \ion{Ne}{x}. However, we observe that the model fits all the absorption lines well that are detected in the spectrum, suggesting that
an absorber with a unique ionization parameter cannot be excluded. In
this scenario we  estimate the location of the ionized
absorbing matter and its thickness.

In the soft state we observe  that the equivalent hydrogen column
density associated with the ionized absorbing matter is
$(6.0^{+0.4}_{-0.8}) \times 10^{23}$ cm$^{-2}$ (uncertainties at 68\%
c.l.).  Using the cosmic abundance for oxygen, neon, and iron reported  by
\cite{Wilms_00} (i.e., $4.9 \times 10^{-4}$, $8.7 \times 10^{-5}$ and
$2.7 \times 10^{-5}$, respectively), we evaluate  the equivalent
column  density for the three elements 
N$_{\rm O} = (2.8 ^{+0.3}_{-0.6}) \times 10^{20} $ cm$^{-2}$,
N$_{\rm Ne} = (5.0 ^{+0.5}_{-1.1}) \times 10^{19} $ cm$^{-2}$ , and
N$_{\rm Fe} = (1.5 ^{+0.2}_{-0.4}) \times 10^{19} $ cm$^{-2}$.
Assuming that the ionized matter is illuminated by 
a power law with spectral index $\Gamma=2$,
we expect for an ionization parameter log$_{\xi} = 4.36$  that the
abundance of \ion{Fe}{xxv} and \ion{Fe}{xxvi} ions with respect to the
neutral iron is $f_{\rm \ion{Fe}{xxv}} \simeq 0.05$ and
$f_{\rm \ion{Fe}{xxvi}} \simeq 0.34$ 
\citep[see][]{Kallman_01}. Using these abundances, we find that the
equivalent column densities associated with \ion{Fe}{xxv} and
\ion{Fe}{xxvi} are
N$_{\rm \ion{Fe}{xxv}} = (7.5 ^{+1.0}_{-2.0}) \times 10^{17} $
cm$^{-2}$ and
N$_{\rm \ion{Fe}{xxvi}} = (5.1 ^{+0.7}_{-1.4}) \times 10^{18} $
cm$^{-2}$, respectively. 

We obtain a relation between the observed equivalent widths of the
absorption lines and the kinetic temperature of the plasma using Eqs. 1-9 shown by \cite{Kotani_00} to estimate the curves of growth of
the observed ions.  For this aim, we adopt the Einstein coefficients
and the values of the oscillator strengths reported by
\cite{Verner_line_96}.
We find that the kinetic temperatures associated with the
\ion{Fe}{xxvi} and \ion{Fe}{xxv} ions are kT$=100^{+45}_{-30}$ keV and
kT$=130^{+470}_{-70}$ keV, respectively. These temperatures
 are compatible
with each other, therefore it is possible to assume that the \ion{Fe}{xxvi} and
\ion{Fe}{xxv} ions are present in the same region of the ionized
absorber and  the kinetic temperature associated with the highly
ionized iron is kT$_{\rm kin}=100^{+45}_{-30}$ keV.

When we assume that the ionized matter is illuminated by a radiation
described as a power law with spectral index $\Gamma=2$, the thermal
temperature of the photoionized matter is
kT$_{\rm th} =7.1^{+0.2}_{-0.1}$ keV for an ionization parameter of
Log$({\xi})_{\rm IA} = 4.37 \pm 0.04$ \citep{Kallman_04}. The relation between
the thermal and kinetic temperature is given by Eq. 1 of
\cite{Yama_01}:
\begin{equation}
\label{eq:yama}
  kT_{\rm th}  +\frac{1}{3}m_{\rm Fe} v^2_{\rm bulk} =  kT_{\rm kin},
\end{equation}
where $m_{\rm Fe}$ is the mass of an iron atom and $v_{\rm bulk}$
denotes the velocity of bulk motion.  \cite{Yama_01} proposed two different
scenarios for large bulk motions corresponding to the kT$_{\rm kin}$
range of 50-500 keV. The first possibility consists of  randomly oriented
turbulent motions in the plasma, and the second attributes the velocity dispersion to the radial velocity gradient. In the second case the ionized absorber may undergo an
outflow, such as a radiation-driven wind. However, we exclude the
second scenario because the observed absorption lines have energies
that are compatible with the rest frame value. Furthermore, \cite{Diaz_16}
suggested that \source could have a mild thermal wind, even if only static
atmospheres have been reported so far.

Using the
values of kT$_{\rm kin}=100^{+45}_{-30}$ keV and
kT$_{\rm th} =7.1^{+0.2}_{-0.1}$ keV, we obtain that the turbulent
velocity is $v_{\rm bulk} =690^{+170}_{-110}$ km s$^{-1}$. This value
is not compatible with the value of 200 km s$^{-1}$
adopted using the model {\sc zxipcf,} and it can explain why the
widths of the absorption lines in the Fe-K region are not well
fit.

A turbulent velocity (associated with the ionized iron) higher
than 500 km s$^{-1}$ was also observed  in the eclipsing NS
binary system AX J1745.6-2901 \citep{Ponti_15}. Using the values shown
in Tab. \ref{tab:line}, we estimate  the flux ratio $R$ between the
K$_{\alpha}$ and K$_{\beta}$ transitions for the \ion{Fe}{xxv} and
\ion{Fe}{xxvi} absorption lines and find $R=1.6^{+0.8}_{-0.6}$ and
$R=1.6 \pm 0.6$, respectively. Figure 4 of
\cite{Risaliti_05} shows that the obtained values of $R$ give a
lower limit  on the turbulent velocity of 100 km
s$^{-1}$   , which is  compatible with a turbulent velocity between
500 and 1000 km s$^{-1}$ for an equivalent column density of neutral
hydrogen of $(6.0_{-1.4}^{+0.9}) \times 10^{23}$ cm$^{-2}$.
  
Assuming that the plasma is in hydrodynamical equilibrium in the
vertical direction to the disk plane, the distance $r_{\rm Fe}$ from
the central source of the absorbing plasma containing   highly
ionized iron  can be estimated from its thermal temperature using
the following expression:
\begin{equation}
\label{eq:equilibrium}
(h/r_{\rm Fe})^2 \times (GM\mu/r_{\rm Fe}) =kT_{\rm th},
\end{equation}
\citep[see][and references therein]{Sidoli_01}, where $M$ is the
NS mass, $\mu\simeq (0.61\;m_H)$ is the mean atomic mass of the
matter assuming that it is fully ionized, $m_H$ is the mass of
an hydrogen atom, $G$ is the gravitational constant, and  $h$ is the scale
height of the absorbing plasma. 
 Assuming that $h/r_{\rm Fe} = \tan\theta=\tan(\pi/2-i)$, where $i$ is the
inclination angle of source, adopting an inclination angle of
$72^{\circ} \pm 3^{\circ}$ and an NS mass of
$1.54\pm 0.22$ M$_{\odot}$, we obtain 
$r_{\rm Fe} = (1.9 \pm 0.7) \times 10^9$ cm.

The electron density $n_e$ in which the iron lines originate can be
estimated using the expression $n_e = L_x/(\xi r_{\rm Fe}^2)$, where
$L_x$ is the luminosity of the source and $\xi$ is the ionization
parameter obtained from the fit. We find that
$n_e = (3.3 \pm 1.3) \times 10^{14}$ cm$^{-3}$, assuming an
uncertainty of 10\% associated with the luminosity $L_x$.  We
estimated the thickness $\Delta r$ of the absorbing plasma using the
expression $N_{\rm H} = n_e \Delta r$, under the assumption that
$n_e \simeq n_H$ and adopting the value of $N_{\rm H} $ obtained from
the best fit (see Tab. \ref{tab:high_flux}). In this way, we obtained
$\Delta r = (1.7_{-0.8}^{+0.7}) \times 10^9$ cm.

\cite{Sidoli_01} suggested that the \ion{O}{viii} and \ion{Ne}{ix}
absorption lines are produced at larger radii than the iron ionized
absorption lines by  invoking a gradient of the
ionization parameter. 
\begin{figure}[ht]
{\hspace{-1.7cm}\includegraphics[width=8.5cm,angle=-90]{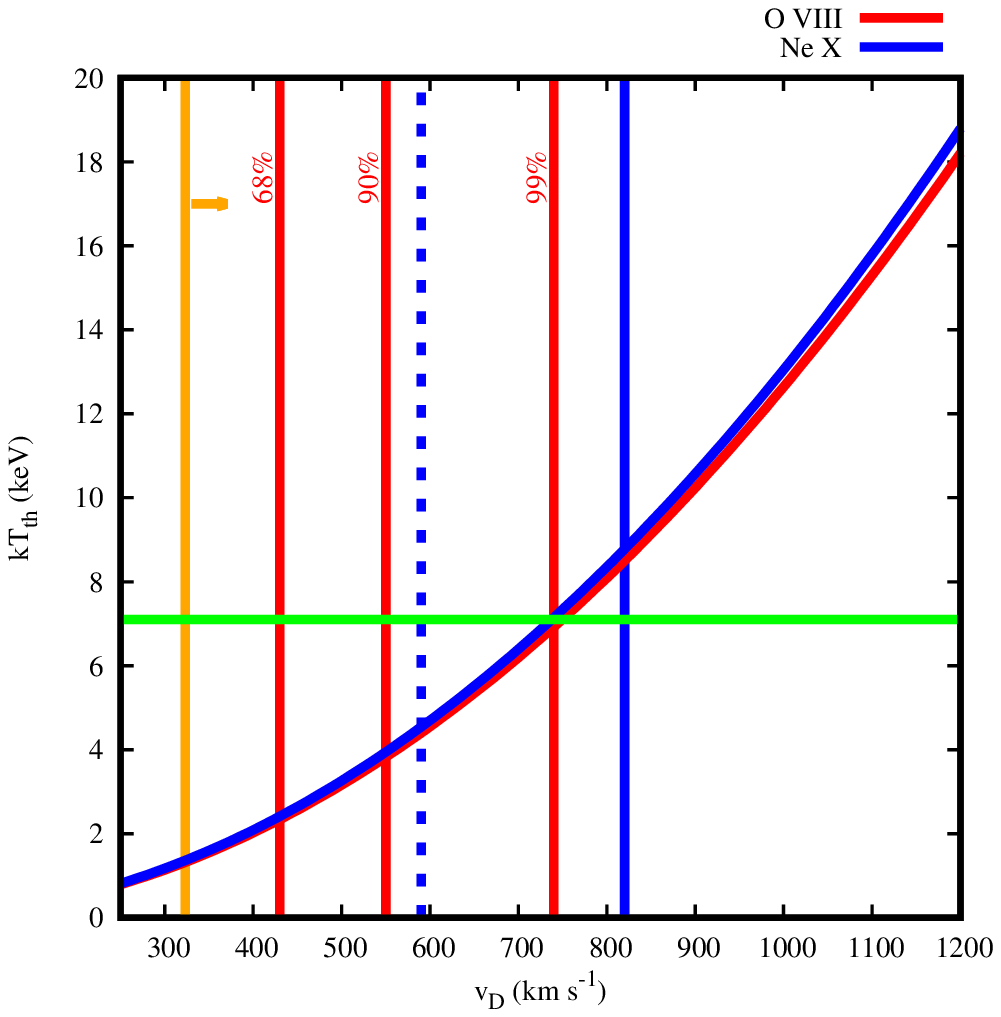}  
}
\caption{Thermal temperature of the ionized absorber as a function
  of the velocity $v_D$ associated with \ion{O}{viii} (red curve) and
  \ion{Ne}{x} (blue curve). The green horizontal line indicates the thermal temperature of
7.1 keV, the solid and dashed blue vertical lines indicate the
measured velocity and the corresponding uncertainties at 68\%
c.l. associated with the \ion{Ne}{x} absorption line.  The red
vertical lines describe  the upper limit   on
the velocity  associated with the \ion{O}{viii} ions at 68\%, 90\%, and
99\% c. l., respectively. The orange vertical line indicates  the velocity $v_D$
associated with  the \ion{O}{viii} ions at the outer radius of the
accretion disc. }
\label{fig:theor}
    \end{figure}
      The obtained velocities, associated with the broadening
    of the lines shown in the seventh column of Tab. \ref{tab:line},  are
    due to both the thermal broadening and the turbulent velocity.
    When the line-of-sight turbulent velocity distribution is described as
    Gaussian, the combined velocity $v_{\rm D}$ is defined as 
\begin{equation}
\label{eq:v}
v^2_{\rm D}=v^2_{\rm th} +v^2_{\rm bulk},  
\end{equation}
where $v_{\rm th}$ and $v_{\rm bulk} $ are the thermal and turbulent velocities, respectively. 
The thermal velocities is expressed as
\begin{equation}
\label{eq:vth}
v^2_{\rm th}= 2 \frac{kT_{\rm th_{I}}}{m_{\rm I}},  
\end{equation}
where kT$_{\rm th_{I}}$ is the thermal energy of the absorber and
$m_{\rm I}$ the mass of the atom that we considered.  Using the
estimated values associated to the ionized iron   and assuming that
the absorber is in hydrodynamical equilibrium along the vertical
direction, we can write
 \begin{equation}
\label{eq:r}
\frac{r_{\rm I}}{r_{\rm Fe}}=\frac{kT_{\rm th_{Fe}}}{kT_{\rm th_{I}}},
\end{equation}
where $r_{\rm Fe}$ and $r_{\rm I}$ are the distances from the NS
where the corresponding lines form. Coronal models tend to have
turbulent velocities that are locally proportional to the virial or
rotational velocity \citep{Woods_96,Iaria_07}, then we can write
\begin{equation} 
\label{eq:vturb}
\frac{r_{\rm I}}{r_{\rm Fe}}=\left(\frac{v_{\rm {bulk}_{Fe}}}{v_{\rm {bulk}_{I}}}\right)^2,
\end{equation}
where $v_{\rm {bulk}_{Fe}}$ and $v_{\rm {bulk}_{I}}$ are the turbulent
velocities of the absorber where the ionized iron lines and the light
ion lines form, respectively. Combining Eqs. \ref{eq:v},
\ref{eq:vth}, \ref{eq:r}, and \ref{eq:vturb}, we obtain a relation
between the thermal energy of the absorber and the observed velocity:
\begin{equation}
\label{eq:fin}
kT_{\rm th_{I}} = \frac{s}{1+s (v^2_{\rm {bulk}_{Fe}}/kT_{\rm th_{Fe}})}v^2_{\rm D} {\;\rm keV},
\end{equation}
where $s = 511/2 \;(m_{\rm I}/m_e) \;1/c^2$ with  $m_e$ the electron mass and $c$ the speed of light.  

Using the best values of the thermal energy and turbulent velocity
associated with the ionized iron (i.e., kT$_{\rm th_{Fe}} \simeq 7.1$
keV and v$_{\rm {bulk}_{Fe}} \simeq 690$ km s$^{-1}$, respectively),
we find the thermal energy as a function of the velocity
$v_{\rm D}$ for the \ion{Ne}{x} absorption line that is shown in
Fig. \ref{fig:theor}.  We expect that if the \ion{Ne}{x}
absorption line is produced in an outer region
with respect to the region where the ionized iron lines are formed,
then the thermal energy of the absorber should be lower than 7.1 keV.
We find that the \ion{Ne}{x} absorption line could be produced in an
absorber with a thermal temperature between 4.5 and 7.1 keV,
corresponding to a value of $v_{\rm D}$ of 590 and 738 km s$^{-1}$,
respectively. Using the equations reported above,    we find that the
\ion{Ne}{x} absorption line forms between
$r = (1.9 \pm 0.7) \times 10^9$ cm and $r = (3.0 \pm 1.1) \times 10^9$
cm, respectively, and  the kinetic energy is between 25.6 and 40 keV. Furthermore,
the spanning range of distances $r$ is compatible with the thickness
of the absorber $\Delta r = (1.7_{-0.8}^{+0.7}) \times 10^9$ cm
estimated for the ionized iron lines,   suggesting that the ionized
iron lines and the \ion{Ne}{x} absorption line could be produced in
the same region of the absorber.
 
Using Eqs. 1-9  of \cite{Kotani_00}, we estimate  that the fraction
$f$ of \ion{Ne}{x} ions compared with the whole population of Ne ions
from the information  that the equivalent column density of neutral neon atoms is
N$_{\rm Ne} = (5.0 ^{+0.5}_{-1.1}) \times 10^{19} $ cm$^{-2}$ and that
the measured equivalent width of the \ion{Ne}{x} absorption line is
$-3.6 \pm 1.0 $ eV.  We obtain   that $f= (3^{+4}_{-2}) \times 10^{-3}$
and $f= (2.5^{+2.5}_{-1.0}) \times 10^{-3}$ for a thermal energy between 
4.5 and 7.1 keV, respectively, corresponding to an equivalent column
density associated with \ion{Ne}{x} ions of $1.5 \times 10^{17}$
cm$^{-2}$.

Regarding the \ion{O}{viii} absorption line, we show the thermal energy
as function of the velocity $v_{\rm D}$ in Fig. \ref{fig:theor}. We find  only an upper limit on the velocity from the fit
(seventh column in Tab. \ref{tab:line}).
 At the 90\% upper
limit, we estimate that $v_D< 550$ km s$^{-1}$ and a thermal energy  
$ <3.82$ keV.  The lower limit on the distance from the NS is
$r_{\rm O}>3.5 \times 10^{9}$ cm. In this case, taking into account that
the geometrical thickness of the absorber is
$\Delta r = (1.7_{-0.8}^{+0.7}) \times 10^9$ cm for a distance from  the
NS of $r = (1.9 \pm 0.7) \times 10^9$ cm, we cannot exclude
the possibility either that the \ion{O}{viii} absorption line forms in the
outer layers of the absorber where the other lines form.  For
$r_{\rm O} =3.5 \times 10^{9}$ cm, we find that the fraction of \ion{O}{viii}
ions with respect the whole population of oxygen atoms is
$f=(2.2^{+1.8}_{-0.7}) \times 10^{-4}$ , which gives an equivalent column
density associated with \ion{O}{viii} ions of $6 \times 10^{16}$
cm$^{-2}$.

Finally, we give an upper limit on the distance, $r$, from the NS
 of the region where the \ion{O}{viii} absorption line  forms  using the relation
$n_e=L_x/(\xi r^2)$ and $ N_H = \Delta r \; /n_e$.  We find that
$\Delta r \simeq 4.4 \times 10^{-10} r^2$ cm. Knowing that the outer
radius of the accretion disk is $r_{\rm disk} = 5 \times 10^{10}$ cm
\citep{Sidoli_01, Iaria_18} and imposing that the thickness of the
absorber is smaller than the radial dimension of the disk,
$\Delta r < r_{\rm disk} $, we obtain that $r \lesssim 1.0 \times 10^{10}$
cm. At this radius the turbulent velocity is 298 km s$^{-1}$, the
thermal velocity is 126 km s$^{-1}$, the velocity $v_D \simeq 323$ km
s$^{-1}$ , and the
kinetic energy is 6.23 keV for the oxygen ions.  For
$r =1.0 \times 10^{10}$ cm we find that the fraction of \ion{O}{viii}
ions with respect to the whole population of oxygen atoms is
$f=(6^{+6}_{-2}) \times 10^{-4}$, which gives an equivalent column
density associated with \ion{O}{viii} ions of $1.7 \times 10^{17}$
cm$^{-2}$.
 
The physical parameters describing the ionized absorber in 
 the soft state of \source are similar to those observed in the
transient eclipsing source AX J1745.6-2901 by \cite{Ponti_15}, who
found   a column density of neutral hydrogen associated with
the absorber higher than $10^{23}$ cm$^{-2}$ and a turbulent velocity
higher than  500  km s$^{-1}$.

 We note that the observed absorption lines we analyzed in the soft spectral state are produced 
within a distance of $4 \times 10^9$ cm from the NS; our result is compatible with the 
distance from the NS of the ionized absorber ($r \leq 1.3 \times 10^{10}$ cm) reported by  \cite{Ponti_19}. Finally, we find that the presence of an ionized absorber   in the hard spectral states is also statistically significant,  unlike the results by 
\cite{Sharma_18}; this is probably related to the different continuum modeling adopted to fit the broadband spectrum.

What is true for the hard state of AX J1745.6-2901 also holds for the hard state
of \source: the ionized iron absorption lines are less prominent, and an
accurate study is not possible.  The different value of the ionization
parameter observed for the \xmm and \nustar
(Log$(\xi)_{IA} = 1.98 \pm 0.05$ and $3.1 \pm 0.3$, respectively) seems to
suggest a rapid ionization of the matter of the absorber, because the
\xmm observation was performed on 2015 September 26 and the \nustar
observation only two days later.  Finally, we observe that the
broadband spectrum of \source is similar to that of AX J1745.6-2901
\citep{Ponti_15}: the spectra in the soft state both show some
features, such as the broad emission line in the soft state or the Compton
hump in the hard state, that can be modeled by a relativistic
reflection component from the accretion disk. Both spectra also show
absorption lines that can be explained by an ionized
absorber.  A similar scenario was also adopted by \cite{Iaria_07} to
describe the Fe-K region of the dipping source X 1624-490, which has an
inclination angle between 60$^{\circ}$ and 70$^{\circ}$. Using Chandra/HETGS data, the authors observed a broad emission line at 6.64 keV
and two narrow absorption lines associated with
\ion{Fe}{xxv} and \ion{Fe}{xxvi} ions. Finally,  the \xmm spectrum of the dipping source  GX 13+1 showed evidence   of a relativistically smeared reflection component with a broad emission line in the Fe-K region and narrow absorption lines associated with highly ionized iron \citep[see][]{diaz_12,Pintore_2014}.

Both \source and AX J1745.6-2901 are binary systems with high
inclination angles, and both sources show eclipses in their light
curves. These similarities suggest that the broad iron lines observed
in the spectra of    high-inclination sources   can be explained as  
relativistically smeared lines instead of   purely Compton broadened lines. The similarities also seem to suggest a similar origin of the broad iron emission line
independently of the inclination angle of the binary system.

\section{Conclusions}
\label{sec:conclusion}
We have analyzed the soft and high state of the transient eclipsing
source \source using \xmm\!\!, \nustar, and \swift spectra taken during
the 1999 and 2015 outburst.
We find that the soft-state continuum emission can be described by a
multicolor  disk blackbody plus a Comptonized component.  The
electron temperature of the Comptonized cloud is between 2 and 4 keV
and the seed-photon temperature is between 0.4 and 0.6 keV.  The inner
temperature of the accretion disk is close to 0.3 keV.  It is
necessary to take into account a smeared reflection component to model
the Fe-K region of the spectrum where a broad emission line is
observed. We find that the width of the emission line cannot be
explained considering the Compton scattering alone, but we have to include
the relativistic smearing in order to obtain a good fit.  The
reflecting region of the accretion disk is between 30 and 2800
gravitational radii, the relative reflection normalization is
$0.22^{+0.12}_{-0.05}$ , and the ionization parameter is
log$(\xi) = 2.80^{+0.20}_{-0.10}$.  Furthermore, an
ionized absorber is observed with an equivalent hydrogen column
density of $6 \times 10^{23}$ cm$^{-2}$ and an ionization parameter
log$({\xi})_{\rm IA}$ between 4.1 and 4.4.  Studying the narrow absorption lines
associated with   \ion{Fe}{xxvi}, \ion{Fe}{xxv},
\ion{Ne}{x,} and \ion{O}{viii}, we obtained information on the
absorber. We find that the absorption lines associated with  highly
ionized iron originate at a distance from the NS of
$(1.9 \pm 0.7) \times 10^9$ cm, the one associated with \ion{Ne}{x} ions
originates between $(1.9 \pm 0.7) \times 10^9$ cm and
$(3.0 \pm 1.1) \times 10^9$ cm, and the line associated with \ion{O}{viii}
ions originates at distances larger than $3.5 \times 10^9$ cm.

In the hard state we do not significantly detect a
thermal component associated with the direct emission of the accretion
disk. The continuum emission was modeled by a Comptonized component
plus a smeared reflection component. We observe an electron
temperature higher than 150 keV and a seed-photon temperature lower
than 0.1 keV.  The reflecting region of the accretion disk is between
6 and 290 gravitational radii, the relative reflection normalization
is $0.48 \pm 0.06,$ and the ionization parameter is
log$(\xi) = 1.99^{+0.05}_{-0.10}$.  In this state we also observed an ionized absorber with an equivalent hydrogen column
density of $1.4 \times 10^{23}$ cm$^{-2}$ and an ionization parameter
log$({\xi})_{\rm IA}$ between 2 and 3.

\section*{Acknowledgements}

This research has made use of data and/or software provided by the High Energy Astrophysics Science Archive Research Center (HEASARC), which is a service of the Astrophysics Science Division at NASA/GSFC and the High Energy Astrophysics Division of the Smithsonian Astrophysical Observatory.\\
The authors acknowledge financial contribution from the agreement
ASI-INAF n.2017-14-H.0, from INAF mainstream (PI: T. Belloni)
and from the HERMES project financed by the Italian Space
Agency (ASI) Agreement n. 2016/13 U.O.
RI and TDS acknowledge the research
grant âiPeskaâ (PI: Andrea Possenti) funded under the INAF
national call Prin-SKA/CTA approved with the Presidential Decree
70/2016. AM acknowledges funding from FSE (Fondo
Sociale Europeo) Sicilia 2020.


\bibliographystyle{aa}                                          
\bibliography{biblio}

\begin{thebibliography}{54}
\expandafter\ifx\csname natexlab\endcsname\relax\def\natexlab#1{#1}\fi

\bibitem[{{Burrows} {et~al.}(2005){Burrows}, {Hill}, {Nousek}, {Kennea},
  {Wells}, {Osborne}, {Abbey}, {Beardmore}, {Mukerjee}, {Short}, {Chincarini},
  {Campana}, {Citterio}, {Moretti}, {Pagani}, {Tagliaferri}, {Giommi},
  {Capalbi}, {Tamburelli}, {Angelini}, {Cusumano}, {Br{\"a}uninger}, {Burkert},
  \& {Hartner}}]{burrows05}
{Burrows}, D.~N., {Hill}, J.~E., {Nousek}, J.~A., {et~al.} 2005, \ssr, 120, 165

\bibitem[{{Cackett} {et~al.}(2010){Cackett}, {Miller}, {Ballantyne}, {Barret},
  {Bhattacharyya}, {Boutelier}, {Miller}, {Strohmayer}, \&
  {Wijnands}}]{Cackett_10}
{Cackett}, E.~M., {Miller}, J.~M., {Ballantyne}, D.~R., {et~al.} 2010, \apj,
  720, 205

\bibitem[{{de Vries} {et~al.}(2003){de Vries}, {den Herder}, {Kaastra},
  {Paerels}, {den Boggende}, \& {Rasmussen}}]{devri_03}
{de Vries}, C.~P., {den Herder}, J.~W., {Kaastra}, J.~S., {et~al.} 2003, \aap,
  404, 959

\bibitem[{{den Herder} {et~al.}(2001){den Herder}, {Brinkman}, {Kahn},
  {Branduardi-Raymont}, {Thomsen}, {Aarts}, {Audard}, {Bixler}, {den Boggende},
  {Cottam}, {Decker}, {Dubbeldam}, {Erd}, {Goulooze}, {G{\"u}del}, {Guttridge},
  {Hailey}, {Janabi}, {Kaastra}, {de Korte}, {van Leeuwen}, {Mauche},
  {McCalden}, {Mewe}, {Naber}, {Paerels}, {Peterson}, {Rasmussen}, {Rees},
  {Sakelliou}, {Sako}, {Spodek}, {Stern}, {Tamura}, {Tandy}, {de Vries},
  {Welch}, \& {Zehnder}}]{herder_01}
{den Herder}, J.~W., {Brinkman}, A.~C., {Kahn}, S.~M., {et~al.} 2001, \aap,
  365, L7

\bibitem[{{di Salvo} {et~al.}(2009){di Salvo}, {D'A{\'{\i}}}, {Iaria},
  {Burderi}, {Dov{\v c}iak}, {Karas}, {Matt}, {Papitto}, {Piraino}, {Riggio},
  {Robba}, \& {Santangelo}}]{Disalvo_09}
{di Salvo}, T., {D'A{\'{\i}}}, A., {Iaria}, R., {et~al.} 2009, \mnras, 398,
  2022

\bibitem[{{Di Salvo} {et~al.}(2015){Di Salvo}, {Iaria}, {Matranga}, {Burderi},
  {D'A{\'{\i}}}, {Egron}, {Papitto}, {Riggio}, {Robba}, \& {Ueda}}]{Disalvo_15}
{Di Salvo}, T., {Iaria}, R., {Matranga}, M., {et~al.} 2015, \mnras, 449, 2794

\bibitem[{{D{\'{\i}}az Trigo} \& {Boirin}(2016)}]{Diaz_16}
{D{\'{\i}}az Trigo}, M. \& {Boirin}, L. 2016, Astronomische Nachrichten, 337,
  368

\bibitem[{{D{\'\i}az Trigo} {et~al.}(2012){D{\'\i}az Trigo}, {Sidoli},
  {Boirin}, \& {Parmar}}]{diaz_12}
{D{\'\i}az Trigo}, M., {Sidoli}, L., {Boirin}, L., \& {Parmar}, A.~N. 2012,
  \aap, 543, A50

\bibitem[{{Done} \& {Gierli{\'n}ski}(2006)}]{Done_06}
{Done}, C. \& {Gierli{\'n}ski}, M. 2006, \mnras, 367, 659

\bibitem[{{Dove} {et~al.}(1997){Dove}, {Wilms}, {Maisack}, \&
  {Begelman}}]{Dove_97}
{Dove}, J.~B., {Wilms}, J., {Maisack}, M., \& {Begelman}, M.~C. 1997, \apj,
  487, 759

\bibitem[{{Egron} {et~al.}(2013){Egron}, {Di Salvo}, {Motta}, {Burderi},
  {Papitto}, {Duro}, {D'A{\`i}}, {Riggio}, {Belloni}, {Iaria}, {Robba},
  {Piraino}, \& {Santangelo}}]{Egron_13}
{Egron}, E., {Di Salvo}, T., {Motta}, S., {et~al.} 2013, \aap, 550, A5

\bibitem[{{Fabian} {et~al.}(1989){Fabian}, {Rees}, {Stella}, \&
  {White}}]{Fabian_89}
{Fabian}, A.~C., {Rees}, M.~J., {Stella}, L., \& {White}, N.~E. 1989, \mnras,
  238, 729

\bibitem[{{Galloway} {et~al.}(2008){Galloway}, {Muno}, {Hartman}, {Psaltis}, \&
  {Chakrabarty}}]{Gallo_08}
{Galloway}, D.~K., {Muno}, M.~P., {Hartman}, J.~M., {Psaltis}, D., \&
  {Chakrabarty}, D. 2008, \apjs, 179, 360

\bibitem[{{Gambino} {et~al.}(2019){Gambino}, {Iaria}, {Di Salvo}, {Mazzola},
  {Marino}, {Burderi}, {Riggio}, {Sanna}, \& {D'Amico}}]{Gambino_19}
{Gambino}, A.~F., {Iaria}, R., {Di Salvo}, T., {et~al.} 2019, \aap, 625, A92

\bibitem[{{Gehrels} {et~al.}(2004){Gehrels}, {Chincarini}, {Giommi}, {Mason},
  {Nousek}, {Wells}, {White}, {Barthelmy}, {Burrows}, {Cominsky}, {Hurley},
  {Marshall}, {M{\'e}sz{\'a}ros}, {Roming}, {Angelini}, {Barbier}, {Belloni},
  {Campana}, {Caraveo}, {Chester}, {Citterio}, {Cline}, {Cropper}, {Cummings},
  {Dean}, {Feigelson}, {Fenimore}, {Frail}, {Fruchter}, {Garmire}, {Gendreau},
  {Ghisellini}, {Greiner}, {Hill}, {Hunsberger}, {Krimm}, {Kulkarni}, {Kumar},
  {Lebrun}, {Lloyd-Ronning}, {Markwardt}, {Mattson}, {Mushotzky}, {Norris},
  {Osborne}, {Paczynski}, {Palmer}, {Park}, {Parsons}, {Paul}, {Rees},
  {Reynolds}, {Rhoads}, {Sasseen}, {Schaefer}, {Short}, {Smale}, {Smith},
  {Stella}, {Tagliaferri}, {Takahashi}, {Tashiro}, {Townsley}, {Tueller},
  {Turner}, {Vietri}, {Voges}, {Ward}, {Willingale}, {Zerbi}, \&
  {Zhang}}]{gehrels04}
{Gehrels}, N., {Chincarini}, G., {Giommi}, P., {et~al.} 2004, \apj, 611, 1005

\bibitem[{{Harrison} {et~al.}(2013){Harrison}, {Craig}, {Christensen},
  {Hailey}, {Zhang}, {Boggs}, {Stern}, {Cook}, {Forster}, {Giommi},
  {Grefenstette}, {Kim}, {Kitaguchi}, {Koglin}, {Madsen}, {Mao}, {Miyasaka},
  {Mori}, {Perri}, {Pivovaroff}, {Puccetti}, {Rana}, {Westergaard}, {Willis},
  {Zoglauer}, {An}, {Bachetti}, {Barri{\`e}re}, {Bellm}, {Bhalerao},
  {Brejnholt}, {Fuerst}, {Liebe}, {Markwardt}, {Nynka}, {Vogel}, {Walton},
  {Wik}, {Alexander}, {Cominsky}, {Hornschemeier}, {Hornstrup}, {Kaspi},
  {Madejski}, {Matt}, {Molendi}, {Smith}, {Tomsick}, {Ajello}, {Ballantyne},
  {Balokovi{\'c}}, {Barret}, {Bauer}, {Blandford}, {Brandt}, {Brenneman},
  {Chiang}, {Chakrabarty}, {Chenevez}, {Comastri}, {Dufour}, {Elvis}, {Fabian},
  {Farrah}, {Fryer}, {Gotthelf}, {Grindlay}, {Helfand}, {Krivonos}, {Meier},
  {Miller}, {Natalucci}, {Ogle}, {Ofek}, {Ptak}, {Reynolds}, {Rigby},
  {Tagliaferri}, {Thorsett}, {Treister}, \& {Urry}}]{harrison_13}
{Harrison}, F.~A., {Craig}, W.~W., {Christensen}, F.~E., {et~al.} 2013, \apj,
  770, 103

\bibitem[{{Iaria} {et~al.}(2009){Iaria}, {D'A{\'{\i}}}, {di Salvo}, {Robba},
  {Riggio}, {Papitto}, \& {Burderi}}]{Iaria_09}
{Iaria}, R., {D'A{\'{\i}}}, A., {di Salvo}, T., {et~al.} 2009, \aap, 505, 1143

\bibitem[{{Iaria} {et~al.}(2016){Iaria}, {Di Salvo}, {Del Santo}, {Pintore},
  {Sanna}, {Papitto}, {Burderi}, {Riggio}, {Gambino}, \& {Matranga}}]{Iaria_16}
{Iaria}, R., {Di Salvo}, T., {Del Santo}, M., {et~al.} 2016, \aap, 596, A21

\bibitem[{{Iaria} {et~al.}(2018){Iaria}, {Gambino}, {Di Salvo}, {Burderi},
  {Matranga}, {Riggio}, {Sanna}, {Scarano}, \& {D'A{\`i}}}]{Iaria_18}
{Iaria}, R., {Gambino}, A.~F., {Di Salvo}, T., {et~al.} 2018, \mnras, 473, 3490

\bibitem[{{Iaria} {et~al.}(2007){Iaria}, {Lavagetto}, {D'A{\'{\i}}}, {di
  Salvo}, \& {Robba}}]{Iaria_07}
{Iaria}, R., {Lavagetto}, G., {D'A{\'{\i}}}, A., {di Salvo}, T., \& {Robba},
  N.~R. 2007, \aap, 463, 289

\bibitem[{{Jain} {et~al.}(2017){Jain}, {Paul}, {Sharma}, {Jaleel}, \&
  {Dutta}}]{Chetana_17}
{Jain}, C., {Paul}, B., {Sharma}, R., {Jaleel}, A., \& {Dutta}, A. 2017,
  \mnras, 468, L118

\bibitem[{{Jansen} {et~al.}(2001){Jansen}, {Lumb}, {Altieri}, {Clavel}, {Ehle},
  {Erd}, {Gabriel}, {Guainazzi}, {Gondoin}, {Much}, {Munoz}, {Santos},
  {Schartel}, {Texier}, \& {Vacanti}}]{jansen_01}
{Jansen}, F., {Lumb}, D., {Altieri}, B., {et~al.} 2001, \aap, 365, L1

\bibitem[{{Kallman} \& {Bautista}(2001)}]{Kallman_01}
{Kallman}, T. \& {Bautista}, M. 2001, \apjs, 133, 221

\bibitem[{{Kallman} {et~al.}(2004){Kallman}, {Palmeri}, {Bautista}, {Mendoza},
  \& {Krolik}}]{Kallman_04}
{Kallman}, T.~R., {Palmeri}, P., {Bautista}, M.~A., {Mendoza}, C., \& {Krolik},
  J.~H. 2004, \apjs, 155, 675

\bibitem[{{Kolehmainen} {et~al.}(2011){Kolehmainen}, {Done}, \& {D{\'{\i}}az
  Trigo}}]{Kole_2011}
{Kolehmainen}, M., {Done}, C., \& {D{\'{\i}}az Trigo}, M. 2011, \mnras, 416,
  311

\bibitem[{{Kotani} {et~al.}(2000){Kotani}, {Ebisawa}, {Dotani}, {Inoue},
  {Nagase}, {Tanaka}, \& {Ueda}}]{Kotani_00}
{Kotani}, T., {Ebisawa}, K., {Dotani}, T., {et~al.} 2000, \apj, 539, 413

\bibitem[{{Mazzola} {et~al.}(2019){Mazzola}, {Iaria}, {Di Salvo}, {Del Santo},
  {Sanna}, {Gambino}, {Riggio}, {Segreto}, {Burderi}, {Santangelo}, \&
  {D'Amico}}]{Mazzola_19}
{Mazzola}, S.~M., {Iaria}, R., {Di Salvo}, T., {et~al.} 2019, \aap, 621, A89

\bibitem[{{Miller} {et~al.}(2013){Miller}, {Parker}, {Fuerst}, {Bachetti},
  {Barret}, {Grefenstette}, {Tendulkar}, {Harrison}, {Boggs}, {Chakrabarty},
  {Christensen}, {Craig}, {Fabian}, {Hailey}, {Natalucci}, {Paerels}, {Rana},
  {Stern}, {Tomsick}, \& {Zhang}}]{Miller_13}
{Miller}, J.~M., {Parker}, M.~L., {Fuerst}, F., {et~al.} 2013, \apjl, 779, L2

\bibitem[{{Miller} {et~al.}(2007){Miller}, {Turner}, {Reeves}, {George},
  {Kraemer}, \& {Wingert}}]{Miller_07}
{Miller}, L., {Turner}, T.~J., {Reeves}, J.~N., {et~al.} 2007, \aap, 463, 131

\bibitem[{{Mitsuda} {et~al.}(1984){Mitsuda}, {Inoue}, {Koyama}, {Makishima},
  {Matsuoka}, {Ogawara}, {Shibazaki}, {Suzuki}, {Tanaka}, \&
  {Hirano}}]{Mitsuda_84}
{Mitsuda}, K., {Inoue}, H., {Koyama}, K., {et~al.} 1984, \pasj, 36, 741

\bibitem[{{Mu{\~n}oz-Darias} {et~al.}(2014){Mu{\~n}oz-Darias}, {Fender},
  {Motta}, \& {Belloni}}]{Munoz_14}
{Mu{\~n}oz-Darias}, T., {Fender}, R.~P., {Motta}, S.~E., \& {Belloni}, T.~M.
  2014, \mnras, 443, 3270

\bibitem[{{{\"O}zel} \& {Freire}(2016)}]{Ozel_16}
{{\"O}zel}, F. \& {Freire}, P. 2016, \araa, 54, 401

\bibitem[{{Pandel} {et~al.}(2008){Pandel}, {Kaaret}, \& {Corbel}}]{Pandel_08}
{Pandel}, D., {Kaaret}, P., \& {Corbel}, S. 2008, \apj, 688, 1288

\bibitem[{{Pintore} {et~al.}(2016){Pintore}, {Sanna}, {Di Salvo}, {Del Santo},
  {Riggio}, {D'A{\`i}}, {Burderi}, {Scarano}, \& {Iaria}}]{Pintore_16}
{Pintore}, F., {Sanna}, A., {Di Salvo}, T., {et~al.} 2016, \mnras, 457, 2988

\bibitem[{{Pintore} {et~al.}(2014){Pintore}, {Sanna}, {Di Salvo}, {Guainazzi},
  {D'A{\`\i}}, {Riggio}, {Burderi}, {Iaria}, \& {Robba}}]{Pintore_2014}
{Pintore}, F., {Sanna}, A., {Di Salvo}, T., {et~al.} 2014, \mnras, 445, 3745

\bibitem[{{Ponti} {et~al.}(2019){Ponti}, {Bianchi}, {De Marco}, {Bahramian},
  {Degenaar}, \& {Heinke}}]{Ponti_19}
{Ponti}, G., {Bianchi}, S., {De Marco}, B., {et~al.} 2019, \mnras

\bibitem[{{Ponti} {et~al.}(2015){Ponti}, {Bianchi}, {Mu{\~n}oz-Darias}, {De
  Marco}, {Dwelly}, {Fender}, {Nandra}, {Rea}, {Mori}, {Haggard}, {Heinke},
  {Degenaar}, {Aramaki}, {Clavel}, {Goldwurm}, {Hailey}, {Israel}, {Morris},
  {Rushton}, \& {Terrier}}]{Ponti_15}
{Ponti}, G., {Bianchi}, S., {Mu{\~n}oz-Darias}, T., {et~al.} 2015, \mnras, 446,
  1536

\bibitem[{{Reeves} {et~al.}(2008){Reeves}, {Done}, {Pounds}, {Terashima},
  {Hayashida}, {Anabuki}, {Uchino}, \& {Turner}}]{Reeves_08}
{Reeves}, J., {Done}, C., {Pounds}, K., {et~al.} 2008, \mnras, 385, L108

\bibitem[{{Reis} {et~al.}(2009){Reis}, {Fabian}, \& {Young}}]{Reis_09}
{Reis}, R.~C., {Fabian}, A.~C., \& {Young}, A.~J. 2009, \mnras, 399, L1

\bibitem[{{Risaliti} {et~al.}(2005){Risaliti}, {Bianchi}, {Matt}, {Baldi},
  {Elvis}, {Fabbiano}, \& {Zezas}}]{Risaliti_05}
{Risaliti}, G., {Bianchi}, S., {Matt}, G., {et~al.} 2005, \apjl, 630, L129

\bibitem[{{Sanna} {et~al.}(2013){Sanna}, {Hiemstra}, {M{\'e}ndez},
  {Altamirano}, {Belloni}, \& {Linares}}]{Sanna_13}
{Sanna}, A., {Hiemstra}, B., {M{\'e}ndez}, M., {et~al.} 2013, \mnras, 432, 1144

\bibitem[{{Sanna} {et~al.}(2017){Sanna}, {Pintore}, {Bozzo}, {Ferrigno},
  {Papitto}, {Riggio}, {Di Salvo}, {Iaria}, {D'A{\`i}}, {Egron}, \&
  {Burderi}}]{Sanna_17}
{Sanna}, A., {Pintore}, F., {Bozzo}, E., {et~al.} 2017, \mnras, 466, 2910

\bibitem[{{Shaposhnikov} {et~al.}(2009){Shaposhnikov}, {Titarchuk}, \&
  {Laurent}}]{Shapo_09}
{Shaposhnikov}, N., {Titarchuk}, L., \& {Laurent}, P. 2009, \apj, 699, 1223

\bibitem[{{Sharma} {et~al.}(2018){Sharma}, {Jaleel}, {Jain}, {Pandey}, {Paul},
  \& {Dutta}}]{Sharma_18}
{Sharma}, R., {Jaleel}, A., {Jain}, C., {et~al.} 2018, \mnras, 481, 5560

\bibitem[{{Shimura} \& {Takahara}(1995)}]{Shimura_1995}
{Shimura}, T. \& {Takahara}, F. 1995, \apj, 445, 780

\bibitem[{{Sidoli} {et~al.}(2001){Sidoli}, {Oosterbroek}, {Parmar}, {Lumb}, \&
  {Erd}}]{Sidoli_01}
{Sidoli}, L., {Oosterbroek}, T., {Parmar}, A.~N., {Lumb}, D., \& {Erd}, C.
  2001, \aap, 379, 540

\bibitem[{{Str{\"u}der} {et~al.}(2001){Str{\"u}der}, {Briel}, {Dennerl},
  {Hartmann}, {Kendziorra}, {Meidinger}, {Pfeffermann}, {Reppin}, {Aschenbach},
  {Bornemann}, {Br{\"a}uninger}, {Burkert}, {Elender}, {Freyberg}, {Haberl},
  {Hartner}, {Heuschmann}, {Hippmann}, {Kastelic}, {Kemmer}, {Kettenring},
  {Kink}, {Krause}, {M{\"u}ller}, {Oppitz}, {Pietsch}, {Popp}, {Predehl},
  {Read}, {Stephan}, {St{\"o}tter}, {Tr{\"u}mper}, {Holl}, {Kemmer}, {Soltau},
  {St{\"o}tter}, {Weber}, {Weichert}, {von Zanthier}, {Carathanassis}, {Lutz},
  {Richter}, {Solc}, {B{\"o}ttcher}, {Kuster}, {Staubert}, {Abbey}, {Holland},
  {Turner}, {Balasini}, {Bignami}, {La Palombara}, {Villa}, {Buttler},
  {Gianini}, {Lain{\'e}}, {Lumb}, \& {Dhez}}]{struder_01}
{Str{\"u}der}, L., {Briel}, U., {Dennerl}, K., {et~al.} 2001, \aap, 365, L18

\bibitem[{{Verner} {et~al.}(1996{\natexlab{a}}){Verner}, {Ferland}, {Korista},
  \& {Yakovlev}}]{Verner_96}
{Verner}, D.~A., {Ferland}, G.~J., {Korista}, K.~T., \& {Yakovlev}, D.~G.
  1996{\natexlab{a}}, \apj, 465, 487

\bibitem[{{Verner} {et~al.}(1996{\natexlab{b}}){Verner}, {Verner}, \&
  {Ferland}}]{Verner_line_96}
{Verner}, D.~A., {Verner}, E.~M., \& {Ferland}, G.~J. 1996{\natexlab{b}},
  Atomic Data and Nuclear Data Tables, 64, 1

\bibitem[{{Wilms} {et~al.}(2000){Wilms}, {Allen}, \& {McCray}}]{Wilms_00}
{Wilms}, J., {Allen}, A., \& {McCray}, R. 2000, \apj, 542, 914

\bibitem[{{Woods} {et~al.}(1996){Woods}, {Klein}, {Castor}, {McKee}, \&
  {Bell}}]{Woods_96}
{Woods}, D.~T., {Klein}, R.~I., {Castor}, J.~I., {McKee}, C.~F., \& {Bell},
  J.~B. 1996, \apj, 461, 767

\bibitem[{{Yamaoka} {et~al.}(2001){Yamaoka}, {Ueda}, {Inoue}, {Nagase},
  {Ebisawa}, {Kotani}, {Tanaka}, \& {Zhang}}]{Yama_01}
{Yamaoka}, K., {Ueda}, Y., {Inoue}, H., {et~al.} 2001, \pasj, 53, 179

\bibitem[{{Zdziarski} {et~al.}(1996){Zdziarski}, {Johnson}, \&
  {Magdziarz}}]{Zidi_96}
{Zdziarski}, A.~A., {Johnson}, W.~N., \& {Magdziarz}, P. 1996, \mnras, 283, 193

\bibitem[{{{\.Z}ycki} {et~al.}(1999){{\.Z}ycki}, {Done}, \& {Smith}}]{Zic_99}
{{\.Z}ycki}, P.~T., {Done}, C., \& {Smith}, D.~A. 1999, \mnras, 309, 561

\end{thebibliography}

\end{document}